\documentclass{aa}
\usepackage{epsfig,graphics}

\begin{document}

\title{Exploring the origin of magnetic fields in massive stars:\\ II.~New magnetic field measurements in cluster and 
field stars.\thanks{Based on observations obtained at the European Southern Observatory,
Paranal, Chile (ESO programmes 087.D-0049(A) and 187.D-0917(A)), and SOFIN observations at the 
2.56\,m Nordic Optical Telescope on La Palma,}}

\author{
S.~Hubrig\inst{1}
\and
M.~Sch\"oller\inst{2}
\and
I.~Ilyin\inst{1}
\and
N.~V.~Kharchenko\inst{3}
\and
L.~M.~Oskinova\inst{4}
\and
N.~Langer\inst{5}
\and
J.~F.~Gonz\'alez\inst{6}
\and
A.~F.~Kholtygin\inst{7}
\and
M.~Briquet\inst{8}\thanks{F.R.S.-FNRS Postdoctoral Researcher, Belgium}
\and
the MAGORI collaboration
}

\institute{
Leibniz-Institut f\"ur Astrophysik Potsdam (AIP), An der Sternwarte~16, 14482~Potsdam, Germany
\and
European Southern Observatory, Karl-Schwarzschild-Str.~2, 85748~Garching, Germany
\and
Main Astronomical Observatory, 27~Academica Zabolotnogo Str., 03680~Kiev, Ukraine
\and
Universit\"at Potsdam, Institut f\"ur Physik und Astronomie, 14476~Potsdam, Germany
\and
Argelander-Institut f\"ur Astronomie, Universit\"at Bonn, Auf dem H\"ugel~71, 53121~Bonn, Germany
\and
Instituto de Ciencias Astronomicas, de la Tierra, y del Espacio (ICATE), 5400~San~Juan, Argentina
\and
Astronomical Institute, Saint-Petersburg State University, Saint-Petersburg, Russia
\and
Institut d'Astrophysique et de G\'eophysique, Universit\'e de Li\`ege, All\'ee du 6 Ao\^ut 17, 4000~Li\`ege, Belgium
}

\date{Received date / Accepted date}

\abstract
{
Theories on the origin of magnetic fields in massive stars
remain poorly developed, because the properties of their magnetic field
as function of stellar parameters could not yet be investigated.
Additional observations are of utmost importance to constrain the conditions 
that are conducive to magnetic fields and to determine 
first trends about their occurrence rate and field strength distribution.
}
{
To investigate whether magnetic fields in massive stars are ubiquitous
or appear only in stars with a specific spectral classification, certain ages, or in a special
environment, 
we acquired 67 new spectropolarimetric observations for 30 massive stars.
Among the observed sample, roughly one third of the stars are probable members of clusters at different ages, 
whereas the remaining stars are field stars not known to belong to any cluster or association.  
}
{
Spectropolarimetric observations were obtained during four different nights using the low-resolution
spectropolarimetric mode of FORS\,2 (FOcal Reducer low dispersion Spectrograph)
mounted on the 8-m Antu telescope of the VLT. Furthermore, we present a number of follow-up observations 
carried out with the high-resolution 
spectropolarimeters SOFIN mounted at the Nordic Optical Telescope (NOT) and HARPS mounted at the ESO 3.6\,m 
between 2008 and 2011. To assess the membership in open clusters
and associations, we used astrometric catalogues with the highest quality
kinematic and photometric data currently available.
}
{
The presence of a magnetic field is confirmed in nine 
stars previously observed
with FORS\,1/2:  
HD\,36879, HD\,47839, CPD$-$28\,2561, CPD$-$47\,2963, HD\,93843, HD\,148937, HD\,149757, HD\,328856, and HD\,164794.
New magnetic field detections at a significance level of at least 3$\sigma$ were achieved in 
five stars: HD\,92206c, HD\,93521, HD\,93632, CPD$-$46\,8221, 
and HD\,157857.
Among the stars with a detected magnetic field, 
five stars belong to open clusters with high membership probability. According 
to previous kinematic studies, five magnetic 
O-type stars in our sample are candidate runaway stars.
}
{}
%{
%}


\titlerunning{Magnetic fields of massive stars}
\authorrunning{S.\ Hubrig et al.}

\maketitle

%________________________________________________________________

%\linenumbers

\section{Introduction}
\label{sect:intro}

During the last years a gradually increasing number of O,
early B-type, and WR stars have been investigated for magnetic
fields, and as a result, about a dozen magnetic O-type and two dozens of early B-type 
stars are presently known (e.g., Donati et al.\ \cite{Donati2006}, 
Hubrig et al.\ \cite{Hubrig2008,Hubrig2011a,Hubrig2011b}, Martins et al.\ \cite{Martins2010}, Wade et al.\ \cite{Wade2011,Wade2012a,
Wade2012b}). 
The recent detections of
magnetic fields in massive stars generate a strong motivation
to study the correlations between evolutionary state, rotation
velocity, and surface composition (Brott et al.\ \cite{Brott2011}, Potter et al.\ \cite{Potter2012}), and 
to understand the origin and the role of magnetic fields in massive stars (Langer \cite{Langer2012}).

The present study is the continuation of our magnetic surveys with FORS\,1/2 at the VLT of a sample of massive O and B-type 
stars in open clusters at different ages and in the field started already in 2005 (e.g., Hubrig et al.\ \cite{Hubrig2008,
Hubrig2011b}).
In 2011 we presented the results of 41 spectropolarimetric FORS\,2 observations of 36 massive stars.
Among the observed sample, roughly half of the stars were probable members of clusters at different 
ages (Hubrig et al.\ \cite{Hubrig2011b}).
The new survey is based on four nights of observations with FORS\,2 
in 2011 May. Furthermore, we present a number of follow-up observations carried out with the high-resolution 
spectropolarimeters SOFIN mounted at the Nordic Optical Telescope (NOT) and HARPS mounted at the ESO 3.6\,m 
between 2008 and 2011. The information on the occurrence rate of magnetic fields
and field strength distribution is critical for answering the principal question of 
the possible origin of magnetic fields in massive stars.

In the following, we present 67 new measurements of magnetic fields in 30 massive stars.
Our observations and the obtained results are described in Sects.~\ref{sect:observations} and
\ref{sect:indivi},
and their discussion is presented in Sect.~\ref{sect:discussion}.

\section{Observations and results}
\label{sect:observations}

\subsection{FORS\,2 observations}
\label{sect:FORS_obs}

\begin{table}
\caption{
List of massive stars observed for this programme.
Spectral classifications are listed according to the Galactic O Star Catalogue 
(Ma{\'{\i}}z-Apell{\'a}niz et al.\ \cite{MaizApellaniz2004}).}
\label{tab:objects}
\centering
\begin{tabular}{llrc}
\hline
\hline
\multicolumn{1}{c}{Name} &
\multicolumn{1}{c}{Other} &
\multicolumn{1}{c}{V} &
\multicolumn{1}{c}{Spectral Type} \\
 &
\multicolumn{1}{c}{Identifier} &
 &
 \\
\hline
HD\,36879      & BD$+$21\,899    & 7.58  & O5 III (f) var      \\ %d
HD\,47839      & 15\,Mon         & 4.64  & O6.5 III (n)(f)    \\ %d
CPD$-$28\,2561 & CD$-$28\,5104   & 10.09 & O6.5 f?p            \\ %d
CPD$-$47\,2963 & CD$-$47\,4551   & 8.45  & O4 III (f)         \\ %d
HD\,77581      & Vela X-1        & 6.93  & B0.5 I ae$^*$      \\
HD\,92206c     & CD$-$57\,3378   & 9.07  & O8.5   V      p   \\
%HD\,92207      & CD$-$58\,3410   & 5.49  & A0 I ae$^*$      \\
HD\,303225     & CD$-$59\,3225   & 9.74  & B$^*$             \\
HD\,93026      & CD$-$58\,3518   & 9.67  & B2 III$^*$        \\
HD\,93521      & BD$+$38\,2179   & 7.04  & O9.5     V      p   \\
HD\,93632      & CD$-$59\,3328   & 8.38  & O5     III    (f) var \\
CPD$-$58\,2611 & CD$-$58\,3526   & 9.55  & O6     V      ((f)) \\      
HD\,93843      & CD$-$55\,5846   & 7.33  & O5 III (f) var      \\ %d
HD\,130298     & CD$-$55\,5846   & 9.29  & O6.5 III (n)(f)    \\ %d
HD\,148937     & CD$-$47\,10855  & 6.77  & O6.5 f?p           \\ %d
HD\,149757     & $\zeta$~Oph     & 2.58  & O9.5   V      nn  \\   
HD\,328856     & CD$-$46\,11016  & 8.50  & O9.5 III$^*$       \\ %d
CPD$-$46\,8221 & CD$-$46\,11017  & 9.19  & O +$^*$             \\ %d
HD\,150958     & CD$-$46\,11019  & 7.29  & O6.5   Ia     (n)f+ \\
HD\,151804     & HR\,6245        & 5.25  & O8 Iaf             \\
HD\,153426     & CD$-$38\,11431  & 7.49  & O9 II-III          \\ %d
HD\,153919     & CD$-$37\,11206  & 6.53  & O6.5 Ia f+         \\ %d
HD\,154643     & CD$-$34\,11503  & 7.15  & O9.5 III           \\ %d
HD\,156041     & CD$-$35\,11432  & 9.2   & B0 V$^*$          \\   
HD\,156154     & CD$-$35\,11445  & 8.04  & O8 Iab (f)         \\ %d
HD\,157857     & BD$-$10\,4493   & 7.81  & O6.5   III    (f) \\   
HD\,164794     & 9\,Sgr          & 5.93  & O4 V ((f))        \\   
HD\,170580     & BD$+$03\,3727   & 6.69  & B2 V$^*$          \\   
HD\,170783     & BD$+$04\,3778   & 7.73  & B5$^*$            \\   
HD\,172175     & BD$-$07\,4642   & 9.44  & O6     I      (n)f \\   
HD\,191612     & BD$+$35\,3995   & 7.84  & O8 f?p           \\ %d
\hline
\end{tabular}
\tablefoot{
All objects were observed with FORS\,2, except
HD\,36879, HD\,47839, and HD\,191612, which were observed with SOFIN,
HD\,151804, which was observed with HARPS, and
HD\,164794, which was observed with HARPS and SOFIN.\\
$^*$~indicates spectral types taken from SIMBAD.
}
\end{table}

\begin{figure*}
\centering
\includegraphics[angle=270,width=0.95\textwidth]{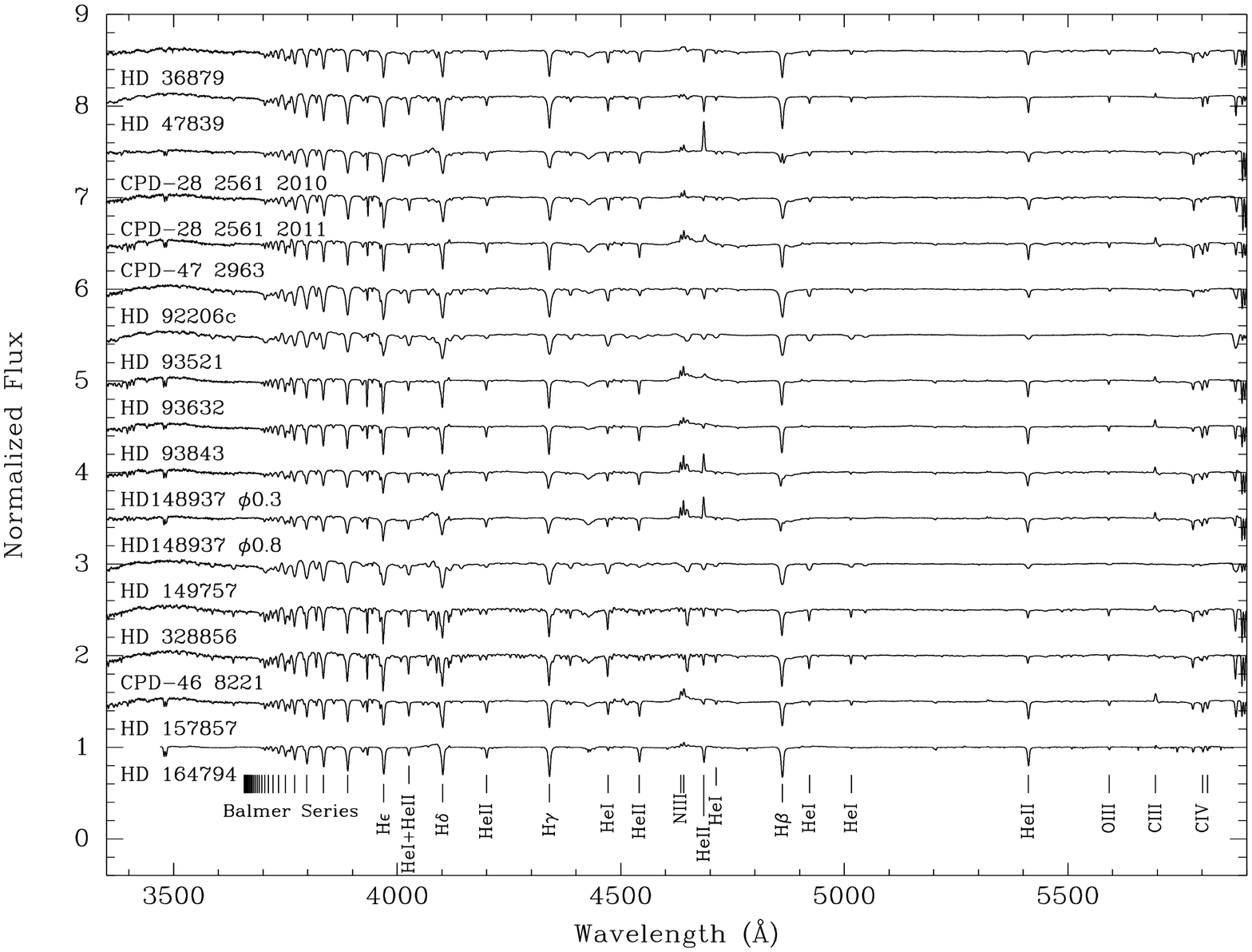}
\caption{
Normalised FORS\,1/2 Stokes $I$ spectra of O-type stars with detected magnetic fields 
observed from 2005 to 2011. Well known spectral lines are indicated;
all Balmer lines from the Balmer jump to H$\beta $ are 
visible. The spectra were offset from 1 by multiples of 0.5 for clarity.}
\label{fig:norm}
\end{figure*}

The spectropolarimetric observations  were carried out
in 2011 May 4--8 in visitor mode
at the European Southern Observatory with FORS\,2 
mounted on the 8-m Antu telescope of the VLT. This multi-mode instrument is equipped with
polarisation analyzing optics, comprising super-achromatic half-wave and quarter-wave 
phase retarder plates, and a Wollaston prism with a beam divergence of 22$\arcsec$  in 
standard resolution mode\footnote{
The spectropolarimetric capabilities of FORS\,1 were moved to
FORS\,2 in 2009.
}.
Polarimetric spectra were obtained with the GRISM~600B and 
the narrowest slit width of 0$\farcs$4 to achieve 
a spectral resolving power of $R\sim2000$. The use of the mosaic detector made of 
blue optimised E2V chips and a  pixel size of 15\,$\mu$m allowed us to cover a large
spectral range, from 3250 to 6215\,\AA{}, which includes all hydrogen Balmer lines 
from H$\beta$ to the Balmer jump.
The spectral types and the visual magnitudes of the studied stars are listed in Table~\ref{tab:objects}.

A detailed description of the assessment of the longitudinal 
magnetic-field measurements using FORS\,2 was presented in our previous papers 
(e.g., Hubrig et al.\ \cite{Hubrig2004a,Hubrig2004b}, and references therein). 
The mean longitudinal 
magnetic field, $\left< B_{\rm z}\right>$, was derived using 

\begin{equation} 
\frac{V}{I} = -\frac{g_{\rm eff} e \lambda^2}{4\pi{}m_ec^2}\ \frac{1}{I}\ 
\frac{{\rm d}I}{{\rm d}\lambda} \left<B_{\rm z}\right>, 
\label{eqn:one} 
\end{equation} 

\noindent 
where $V$ is the Stokes parameter that measures the circular polarisation, $I$ 
is the intensity in the unpolarised spectrum, $g_{\rm eff}$ is the effective 
Land\'e factor, $e$ is the electron charge, $\lambda$ is the wavelength, $m_e$ the 
electron mass, $c$ the speed of light, ${{\rm d}I/{\rm d}\lambda}$ is the 
derivative of Stokes $I$, and $\left<B_{\rm z}\right>$ is the mean longitudinal magnetic 
field. 
Errors are determined in the original spectra from photon statistics and
propagated to the resulting $V/I$ spectra.
The error of $\left<B_{\rm z}\right>$ is obtained from the formal error
of the error-weighted least-squares fit.
%Errors on ${{\rm d}I/{\rm d}\lambda}$ or $1/I$ are neglected.

Longitudinal magnetic fields were measured in two ways: using only the absorption hydrogen Balmer 
lines or the entire spectrum including all available absorption lines.
The lines that show evidence of emission were not used to determine the magnetic field strength. 
In addition, a rectification was carried out for a few massive stars with continuum slopes detected in their 
Stokes $V$ spectra.
The feasibility of longitudinal magnetic field measurements in massive stars 
using low-resolution spectropolarimetric observations has been demonstrated
by previous studies of O and B-type stars
(e.g., Hubrig et al.\ \cite{Hubrig2006,Hubrig2008,Hubrig2009a,Hubrig2011a,Hubrig2011b}).

\begin{table*}
\caption{
Longitudinal magnetic fields measured with FORS\,2 in the studied sample.
Entries related to previous measurements are indicated by an asterisk in the second column of the Table. 
The wavelength region used for the magnetic field measurements is indicated in the sixth column. The seventh column provides 
the rotation phases for the Of?p star HD\,148937, which has a known rotation period of about seven days (Naz\'e et al.\ \cite{Naze2010}). 
In the last column ND stands for new detections,
PD for previous detections (Hubrig et al.\ \cite{Hubrig2008,Hubrig2011b}),
and CD for confirmed detections.
All quoted errors are 1$\sigma$ uncertainties.
}
\label{tab:fields}
\centering
\begin{tabular}{lcrr @{$\pm$} lr @{$\pm$} lccc}
\hline
\hline
\multicolumn{1}{c}{Name} &
\multicolumn{1}{c}{MJD} &
\multicolumn{1}{c}{S/N} &
\multicolumn{2}{c}{$\left< B_{\rm z}\right>_{\rm all}$} &
\multicolumn{2}{c}{$\left< B_{\rm z}\right>_{\rm hydr}$} &
\multicolumn{1}{c}{Spectral} &
\multicolumn{1}{c}{Phase} &
\multicolumn{1}{c}{Comment} \\
 &
 &
 &
\multicolumn{2}{c}{[G]} &
\multicolumn{2}{c}{[G]} &
\multicolumn{1}{c}{Regions (\AA)} &
 &
 \\
\hline
CPD$-$28\,2561 & $^{*}$55338.969 &2262  & $-$381 & 122 &  $-$534 & 167 &3580--4588 & & PD \\ % d /work2/shubrig/data_O_5May11; 
                   & 55685.982& 2481    &  99 &  82 &     65 & 108 &   &\\ % /work1/shubrig/home3/shubrig/grism600B_may2011/all
                   & 55686.984 & 2516   & $-$44 &  80 &       6 & 102 &    &  \\
                   & 55687.980 & 2490   &   269 &  81 &     281 &  90 & & & CD   \\

CPD$-$47\,2963  & $^{*}$55337.094 & 2722 &  $-$190 &  62 &  $-$154 & 96 &3696--5622 & & PD \\ % d
                   & 55686.019 & 2691   & $-$66 &  57 &  $-$119 & 82 &  & &\\
                   & 55687.037 & 3035  & $-$169 & 52 &   $-$77 & 76 & & & CD \\
                   & 55688.013 & 2164   &  95 &  69 &     148 &101 &  & &\\

Vela\,X-1         & 55686.096 & 3604    & $-$15 &  30 &   $-$11 &  45 &3562--5886  & &\\ % d
                  & 55687.058 & 2894    & $-$80 &  32 &  $-$114 &  50 &  &&\\
                  & 55688.048 & 2433    &  57 &  37 &      88 &  69 & &&\\

HD\,92206c        & 55686.135 & 2393    & 204 &  46 &     140 &  88 &3658--5944 & & ND \\ % d
                  & 55687.085 & 3439    & $-$104 &  30 &  $-$123 &  54 &  & &  \\
                  & 55688.151 & 2528    &  10 &  43 &      23 &  46 & &&\\
%HD\,92207new      & 55688.168 &    & $-$333 &  41 &  $-$338 &  50 & & & ND \\

HD\,303225        & 55688.220 & 1902    & $-$59 &  54 &   $-$81 & 90 &3447--5299 && \\ % d

HD\,93026         & 55687.247 & 1647    & $-$19 &  75 &   $-$52 & 100 &3470--4900 & &\\ % d

HD\,93521         & 55686.042 & 3891    &  32 &  52 &      45 &  73 &3689--5414 &&\\  % d
                  & 55687.015 & 3304   & $-$202 &  66 &  $-$293 &  91 & & & ND \\
                  & 55688.063 & 3059    & $-$74 &  62 &   $-$64 &  68 & &&\\

HD\,93632         & 55687.217 & 2927    & $-$152 &  50 & $-$120 &  62 &3318--4593 & & ND\\ % d
                  & 55688.198 & 2443    & $-$10 &  32 &      5 &  68 & &&\\

CPD$-$58\,2611     & 55687.151 & 2631   & $-$58 &  44 &   $-$80 &  59 &3580--4588  &&\\% d
                   & 55688.125 & 2005   & $-$25 &  56 &    $-$9 &  74 &  &&\\

HD\,93843     & $^{*}$55339.099 & 3504  & $-$157 &  42 &  $-$173 &  56 & 3370--5545 & & PD \\ % d
                  & 55686.115 & 3230    & $-$34 &  38 &   $-$12 &  52 &   &&\\
                  & 55687.185 & 3385    & $-$132 &  39 &   $-$93 &  56 & & & CD   \\
                  & 55688.180 & 2647    & $-$63 &  47 &   $-$130&  61 & &   &\\
HD\,130298      & $^{*}$55339.159 &3037 &     113 &  38 &     193 &  62 &3573--5284 & & PD \\ % d
                & 55686.347 &  2264     & $-$64 &  66 &   $-$64 &  85 &  &&\\
                  & 55688.261 & 2699    &  59 &  55 &     122 &  72 &  &&\\

HD\,148937 & $^{*}$54550.416 & 2616     & $-$276&  88 &  $-$145 & 104 &3210--3599 &0.631 & PD \\ % d
%               & 55336.307 &       & $-$297 &  62 &  $-$293 &  85 & & & PD \\ % bad, see Schoeller et al.
               & $^{*}$55337.285 & 3364 & $-$204 &  71 &  $-$225 & 103 & & 0.529 &  \\
               & $^{*}$55339.206 & 2253 & $-$290 &  85 &  $-$389 & 129 & & 0.802 & PD \\
             & 55686.370 &    2664       &  $-$110 &  90 &  $-$165 & 115 & & 0.172 & CD \\
               & 55687.265 & 3083       & $-$87 &  74 &  $-$137 &  95 & & 0.299 & CD \\
               & 55688.285 & 2941      & $-$139 &  33 &  $-$174 & 121 & & 0.444 & CD \\
               & 55689.361 & 2522      & $-$241 &  91 &  $-$229 & 118 & & 0.597 & CD \\

\hline
\end{tabular}
\end{table*}

\addtocounter{table}{-1}
\begin{table*}
\caption{
Continued.
}
\centering
\begin{tabular}{lcrr @{$\pm$} lr @{$\pm$} lccc}
\hline
\hline
\multicolumn{1}{c}{Name} &
\multicolumn{1}{c}{MJD} &
\multicolumn{1}{c}{S/N} &
\multicolumn{2}{c}{$\left< B_{\rm z}\right>_{\rm all}$} &
\multicolumn{2}{c}{$\left< B_{\rm z}\right>_{\rm hydr}$} &
\multicolumn{1}{c}{Spectral} &
\multicolumn{1}{c}{Phase} &
\multicolumn{1}{c}{Comment} \\
 &
 &
 &
\multicolumn{2}{c}{[G]} &
\multicolumn{2}{c}{[G]} &
\multicolumn{1}{c}{Regions} &
 &
 \\
\hline
$\zeta$~Oph & $^{*}$54609.343 & 4083 & 105   &  45  &  54  &  54 &3645--4960 & & PD \\ % d
                & 55686.321 & 3719   & $-$164 &  62 &  $-$218 & 70  & & & CD \\
               & 55687.116 & 3239    & $-$83  &  63 &  $-$90  & 81  & &    &  \\
               & 55687.287 & 2659    & $-$153 &  75 &  $-$222 & 97  &  &   &  \\
               & 55687.385 & 2592    & $-$101 &  88 &   $-$84 & 104 &  &   &  \\
               & 55688.103 & 1964    &  18 & 113 &      18 & 144 &  &   &  \\
               & 55688.240 & 2792    &  95 &  68 &      97 &  80 &  &   &  \\
               & 55688.335 & 2226    & $-$72 &  94 &   $-$37 & 108 &  &   &  \\
               & 55688.433 & 2486    & $-$16 &  89 &      14 & 110 &  &   &  \\
               & 55689.349 & 2193    &  58 &  90 &      52 & 117 & &    &  \\
HD\,328856 & $^{*}$55336.370 & 3239  & $-$173 &  53 &  $-$155 &  65 &3589--5669 & & PD \\ % d
             & $^{*}$55339.223 &2975 &  $-$149 &  48 &   $-$75 &  72 & & & PD \\
               & 55686.417 & 2309     & $-$43 &  48 &   $-$96 &  70 & &   & \\ % Hogg 22 
               & 55687.351 & 2612    &   9 &  46 &      21 &  64 & & & \\
               & 55688.301 & 2279     & $-$128 &  50 &  $-$250 &  77 & & & CD \\
CPD$-$46\,8221 & 55686.393 & 2389    & $-$132 &  42 &   $-$65 &  63 & 3569--5158& & ND \\ % Hogg 22 d
               & 55688.319 & 2195    & $-$83 &  45 &   $-$69 &  68 &  &  &  \\ % Hogg 22
HD\,150958    & 55686.431 & 1957     & 133 &  65 &     122 & 118 & 3695--4974  &  &  \\ % Hogg 22 d
               & 55687.372 & 2206    & $-$105 &  63 &   $-$74 & 106 &  &  &  \\ % Hogg 22
HD\,153426   & $^{*}$55336.338 & 3446 &   $-$27 &  53 &   $-$10 &  62&3650--4874 & &\\ % d
             & $^{*}$55339.246 &3136 &  $-$171 &  55 &  $-$275 &  70 & & & PD \\
            & 55687.434 & 2318       & 19  &  65 &      62 &  82 &  & &\\ 
HD\,153919  & $^{*}$55337.341 &2727  & $-$213 &  68 &  $-$119 &  95 &3680--4452 & & PD \\ % d
           & 55687.421 & 2727        & 26 &  82 &     152 &  102 &  &&\\

HD\,154643 & $^{*}$55339.340 & 3382  &  110 &  34 &     121 &  52 &3376--4910 & & PD \\ % d
              & 55689.374 & 2925     &  73 &  42 &     163 &  58&  & &\\

HD\,156041     & 55689.393 &1501     &  71 &  69 &     138 &  109 &3689--5663  & & \\ % d

HD\,156154 & $^{*}$55337.358 & 2753  & $-$118 &  38 &  $-$167 &  54 &3617--4974 & & PD \\ % d
               & 55688.418 & 2498    &  19 &  54 &     117 &  72 &  & &\\

HD\,157857     & 55687.401 &  3035   & $-$110 &  46 &  $-$226 &  61 &3383--4974 & & ND \\ % d
               & 55688.348 & 2785    & $-$116 &  50 &   $-$51 &  72 & &  &\\
HD\,170580     & 55688.403 & 2579    & $-$55 &  31 &   $-$77 &  47 &3343--6196 &  &\\ % d
HD\,170783     & 55688.389 & 3016    & $-$19 &  34 &    $-$2 &  43 &3326--5885 &  &\\ % d
HD\,172175     & 55688.370 & 2210    & $-$73 &  94 &   $-$79 &  124 &3670--4480 &  &\\ % d
\hline
\end{tabular}
\end{table*}

Ten targets were observed once, 13 were observed two or three times.
HD\,148937 was observed four times to assess the magnetic field variability over the rotation cycle.
Apart from this star, which has a rotation period of seven days (Naz\'e et al.\ \cite{Naze2010}), 
no exact rotation periods are known for the other stars in our sample. The star $\zeta$~Oph was observed nine times
with the aim to search for magnetic/rotational periodicity (see Sect.~\ref{sect:indivi}). 

The results of our magnetic field measurements are presented in Table~\ref{tab:fields}.
In the first two columns, we provide the star names and the modified Julian dates at the middle of 
the exposures, followed by the signal-to-noise ratio in the individual spectra in Col.~3.
In Cols.~4 and 5, we present the longitudinal magnetic 
field $\left<B_{\rm z}\right>_{\rm all}$ using the whole spectrum, and the longitudinal magnetic field 
$\left<B_{\rm z}\right>_{\rm hyd}$ using only the hydrogen lines. 
All quoted errors are 1$\sigma$ uncertainties.
In Col.~6 we indicate the wavelength regions used for the magnetic field measurements. The spectral regions containing
the interstellar lines of the \ion{Ca}{ii} doublet are usually not used in the measurements.
In Col.~7 we list 
the rotation phases for the Of?p star HD\,148937, which is the only star in our FORS\,2 observations with a known period of
about seven days.
% (Naz\'e et al.\ \cite{Naze2010}).
Finally, in the last column we identify new detections by ND, previous detections by PD and 
confirmed detections by CD.

In our sample of 25 stars observed with FORS\,2 in 2011 May the presence of a magnetic field is confirmed in six 
stars previously observed
with FORS\,1/2 (Hubrig et al.\ \cite{Hubrig2008,Hubrig2011b,Hubrig2011c}):  
CPD$-$28\,2561, CPD$-$47\,2963, HD\,93843, HD\,148937, $\zeta$~Oph, and HD\,328856.
New magnetic field detections were achieved in five stars: HD\,92206c, HD\,93521, HD\,93632, CPD$-$46\,8221, 
and HD\,157857.
Although previous, mostly single observations of the five O-type stars HD\,130298, HD\,153426, HD\,153919, HD\,154643, 
and HD\,156154 indicated the possible presence of weak magnetic fields (Hubrig et al.\ \cite{Hubrig2011b}),
our recent observations do not reveal magnetic fields
at a significance level of 3$\sigma$ in these stars. We note that these recent observations 
have been carried out on single epochs for four out of five stars, only  HD\,130298 was observed on two different nights.
Since the rotation periods for them are unknown, the non-detection of their magnetic fields can probably 
be explained by the strong dependence of the longitudinal magnetic field on the rotational aspect.

The strongest magnetic fields are detected in the two Of?p stars CPD$-$28\,2561 and  HD\,148937.
Walborn (\cite{Walborn1973})
introduced the Of?p category for massive O stars displaying recurrent spectral variations in 
certain spectral lines, sharp emission or P~Cygni profiles in \ion{He}{i} and the Balmer lines, and strong \ion{C}{iii} 
emission lines around 4650\,\AA{}.
Only five Galactic Of?p stars are presently known: HD\,108, NGC\,1624-2, CPD$-$28\,2561, HD\,148937, and HD\,191612 
(Walborn et al.\ \cite{Walborn2010}), and all of them show evidence for the presence of magnetic fields (Martin et al.\ \cite{Martins2010},
Wade et al.\ \cite{Wade2012a}, Hubrig et al.\ \cite{Hubrig2008,Hubrig2011b}, Donati et al.\ \cite{Donati2006}).
The record holder is the faintest star in the sample of Of?p stars,  NGC\,1624-2, for which a longitudinal magnetic field
of the order of 5\,kG was recently detected by Wade et al.\ (\cite{Wade2012a}). 

Besides the small group of Of?p stars, our measurements indicate that the presence of magnetic fields can be expected 
in stars of different classification
categories and at different evolutionary stages, from young main-sequence stars up to the supergiant stages.
The different spectral appearance of stars with confirmed and new detections is illustrated in Fig.~\ref{fig:norm} 
where we present 
normalised FORS\,1/2 spectra in integral light. Among the whole sample, the two Of?p stars CPD$-$28\,2561 and  HD\,148937
exhibit the most conspicuous spectral variability clearly visible in Fig.~\ref{fig:norm}, showing the spectra of these stars 
obtained at two different epochs. To calculate the rotation phase of the observed spectra of HD\,148937, we used the 
rotation period of 7.03\,d
determined by Naz\'e et al.\ (\cite{Naze2008}). 
The stars with confirmed and new magnetic field detections are discussed in more detail in Sect.\ \ref{sect:indivi}.

\subsection{SOFIN and HARPS observations}
\label{subsect:sofin}

Four O-type stars, HD\,36879, 15\,Mon, 9\,Sgr, and HD\,191612, previously observed with FORS\,1 and 
SOFIN (Hubrig et al.\ \cite{Hubrig2008,Hubrig2009a,Hubrig2010}), 
were followed-up by nine additional SOFIN spectropolarimetric observations from 2008 September and 2010 July.
Spectral classifications of these stars are presented in Table~\ref{tab:objects}. Among these stars, only
for the Of?p star HD\,191612 was the presence of a magnetic field of the order
of a few hundred Gauss previously reported by Donati et al.\ (\cite{Donati2006}).
We used the low-resolution camera
($R\approx30\,000$)
of the echelle spectrograph SOFIN (Tuominen et al.\ \cite{Tuominen1999})
mounted at the Cassegrain focus of the Nordic Optical Telescope (NOT).
With the 2K Loral CCD detector we registered 40 echelle orders partially covering
the range from 3500 to 10\,000\,\AA{} 
with a length of the spectral orders of about 140\,\AA{} at 5500\,\AA{}.
The polarimeter is located in front of the entrance slit of the spectrograph and consists of a fixed 
calcite beam splitter aligned along the slit and a rotating super-achromatic quarter-wave plate. Two spectra 
polarised in opposite sense are recorded simultaneously for each echelle order providing sufficient separation 
by the cross-dispersion prism.
Two to four sub-exposures with the quarter-wave plate angles separated 
by $90^\circ$ were used to derive circularly polarised spectra.
A detailed description of the SOFIN spectropolarimeter and its polarimetric data reduction is given in
Ilyin (\cite{Ilyin2012}).

The spectra were reduced with the 4A software package (Ilyin \cite{Ilyin2000}). 
Bias subtraction, master flat-field correction, 
scattered light subtraction, and weighted extraction of spectral orders comprise the standard steps 
of spectrum processing.
A ThAr spectral lamp is used for wavelength calibration, taken before
and after each target exposure to minimise temporal variations in the spectrograph. The typical 
signal-to-noise
ratio (S/N) for observations of 9\,Sgr is between 280 and 380, while for HD\,36879 and HD\,191612 it is about 200.
The highest S/N ratio of 440 was achieved for 15\,Mon.   

\begin{table}
\caption[]{
Magnetic field measurements of five O-type stars using SOFIN and HARPS observations. 
Already published measurements
are marked by asterisks.
All quoted errors are 1$\sigma$ uncertainties.
}
\label{tab:log_meassofin}
\centering
\begin{tabular}{rrr@{$\pm$}lr@{$\pm$}lc}
\hline \hline\\[-7pt]
\multicolumn{1}{c}{MJD/HJD} &
\multicolumn{1}{c}{Phase} &
\multicolumn{2}{c}{$\left<B_{\rm z}\right>_{\rm all}$ [G]} &
\multicolumn{2}{c}{$\left<B_{\rm z}\right>_{\rm n}$ [G]} &
\multicolumn{1}{c}{Instr.} \\
\hline\\[-7pt]
\multicolumn{7}{c}{HD\,36879} \\
\hline\\[-7pt]
$^{*}$54345.389 & & 180  &52  &  \multicolumn{2}{c}{}  & FORS\,2 \\
2454723.737 &   & $-$125 & 78 & 53& 75 & SOFIN \\
2455195.615 &   & $-$276 & 89 & $-$64& 92 & SOFIN \\
\hline\\[-7pt]
\multicolumn{7}{c}{15\,Mon}\\
\hline\\[-7pt]
  $^{*}$54609.968 & & 134  &52  &  \multicolumn{2}{c}{} & FORS\,2 \\
  2455201.553 & &  $-$162 & 37 & $-$62 & 42 & SOFIN \\
\hline\\[-7pt]
\multicolumn{7}{c}{HD\,151804}\\
\hline\\[-7pt]
$^{*}$53476.369 & & $-$151  &90  &  \multicolumn{2}{c}{} & FORS\,2 \\
$^{*}$53571.025 & & 68  &65  &  \multicolumn{2}{c}{} & FORS\,2 \\
$^{*}$53596.096 & &82  &46  &  \multicolumn{2}{c}{} & FORS\,2 \\
55708.356& & $-$104& 38 &  $-$31 & 36 & HARPS \\
\hline\\[-7pt]
\multicolumn{7}{c}{9\,Sgr}\\
\hline\\[-7pt]
$^{*}$53520.357 & & $-$114  &66  &  \multicolumn{2}{c}{} & FORS\,2 \\
$^{*}$53594.119 & & 211  &57  &  \multicolumn{2}{c}{} & FORS\,2 \\
$^{*}$53595.096 & &$-$165  &75  &  \multicolumn{2}{c}{} & FORS\,2 \\		
2455074.387 & &  187  & 56 &  $-$33 & 60 & SOFIN \\
2455075.384  & &  242 &  74 & 72&  76 & SOFIN \\
2455078.396  & &  238 & 59 & 66 & 66 & SOFIN \\
2455081.382  & &  $-$265 & 65 & $-$42& 62 & SOFIN \\
55707.323 & & 210& 42 & 39 & 41 & HARPS \\
\hline\\[-7pt]
\multicolumn{7}{c}{HD\,191612}\\
\hline\\[-7pt]
  $^{*}$2454721.347 & 0.43 &  357 & 58 &   67 & 65 & SOFIN \\
  2455395.592      & 0.68  & $-$ 144 & 34 & $-$39 & 36 & SOFIN \\
  2455407.583       & 0.68 & $-$154 & 63 & $-$67& 70 & SOFIN \\
\hline\\[-7pt]
\hline
\end{tabular}
\end{table}

In addition, the two O-type stars, HD\,151804 and 9\,Sgr, previously analyzed using FORS\,1 data 
(Hubrig et al.\ 
\cite{Hubrig2008}) were observed on 2011 May 25 and 26 with a S/N of about 280 with the HARPSpol polarimeter
(Snik et al.\ \cite{Snik2011}) feeding the HARPS spectrometer 
(Mayor et al.\ \cite{Mayor2003}) at the ESO 3.6\,m telescope on La Silla within the framework of the program
187.D-0917(A). 
This HARPS spectropolarimetric material
became recently publically available in the ESO archive. The HARPS archive spectra cover the wavelength 
range 3780--6913\,\AA{}, with a small gap around 5300\,\AA{}. 
Each HARPSpol observation is usually split into four to eight sub-exposures, 
obtained with four different orientations of the quarter-wave retarder plate relative to the 
beam splitter of the circular polarimeter. The reduction was performed using the HARPS data reduction 
software available at the ESO headquarters in Germany. 
The Stokes~$I$ and $V$ parameters were derived following the ratio method described by 
Donati et al.\ (\cite{Donati1997}), 
ensuring in particular that all spurious signatures are removed at first order. 
Null polarisation spectra (labeled with $n$ in Table~\ref{tab:log_meassofin}) were calculated by combining the 
sub-exposures 
in such a way that the polarisation cancels out, allowing us to verify that no spurious signals are present in the data.

HARPS and SOFIN spectra have been measured using the moment technique previously introduced by Mathys who 
discussed at length the diagnostic potential of high-resolution circularly polarised spectra using this technique 
in numerous papers (e.g.,\ Mathys \cite{Mathys1993, Mathys1995a, Mathys1995b}).
Wavelength shifts between right- and left-hand side circularly polarised spectra are
interpreted in terms of a longitudinal magnetic field $\left<B_{\rm z}\right>$.
$\left<B_{\rm z}\right>$ is obtained by performing a least-squares fit,
forced through the origin, of the wavelength shifts
$\lambda_R-\lambda_L$, as a function of $2\overline{g}\Delta\lambda_{\rm z}$.
In this process, the individual measurements are weighted by the respective
$1/\sigma^2(\lambda_R-\lambda_L)$.
The uncertainty of the derived value of the longitudinal magnetic field
is estimated by its standard error $\sigma(\left<B_{\rm z}\right>)$,
derived from the least-squares analysis
(e.g., Mathys \& Hubrig \cite{MathysHubrig1995}).
The results of the longitudinal magnetic field measurements using mostly \ion{He}{i} and  \ion{He}{ii} lines
are presented in Table~\ref{tab:log_meassofin}.

\begin{figure}
\centering
\includegraphics[angle=270,width=0.45\textwidth]{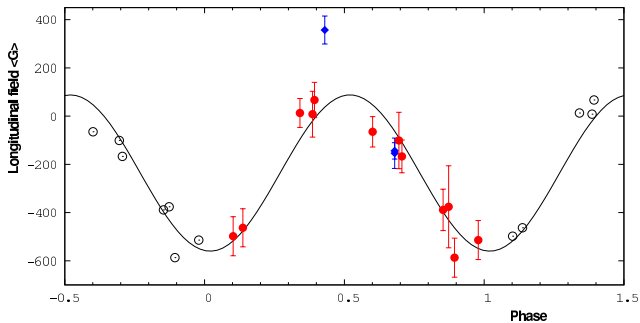}
\caption{
 Longitudinal magnetic field variation of the Of?p star HD\,191612 according to the 537.6\,d period 
determined by Howarth et al.\ (\cite{Howarth2007}).
Red symbols correspond to ESPaDOnS observations, while blue symbols are 
our SOFIN measurements (Hubrig et al.\ \cite{Hubrig2010} and Table~\ref{tab:log_meassofin}).
Note that the measurement errors for both ESPaDOnS and SOFIN observations are of similar order. 
}
\label{fig:ofp191}
\end{figure}

The high-resolution spectropolarimetric observations confirm our previous detections of magnetic fields in the stars HD\,36879
and 9\,Sgr. The magnetic nature of the slowly rotating Of?p star HD\,191612 ($P=538$\,d) was for the 
first time reported by Donati et al.\ (\cite{Donati2006}) who observed it over four nights in 2005 June using 
the ESPaDOnS spectropolarimeter installed on the 3.6-m Canada-France-Hawaii Telescope and announced the detection 
of a mean longitudinal 
field $\langle B_{\rm z}\rangle=-220\pm38$\,G. Hubrig et al.\ (\cite{Hubrig2010}) reported on a single 
measurement of 
the magnetic field of HD\,191612, $\langle B_{\rm z}\rangle=450\pm153$\,G, using SOFIN observations acquired in 
2008 September at rotational phase 0.43.
Wade et al.\ (\cite{Wade2011}) presented 13 new field measurements of this star demonstrating 
that the magnetic data can be modeled as a periodic, sinusoidal signal with a period of 538\,d inferred from spectroscopy.
The authors mention in their work that
the SOFIN observation of the longitudinal field  of HD\,191612, $\langle B_{\rm z}\rangle=450\pm153$\,G,
does not agree very well with the observed field variation, which corresponds
to (an essentially) consistently negative longitudinal magnetic field. To check the consistency of our measurements we 
re-analyzed the previous 
SOFIN observations using an extended list of blend free spectral lines with known atomic parameters. The obtained results,
$\langle B_{\rm z}\rangle=357\pm58$\,G, indicates, however, that contrary to the results of Wade et al.\ 
(\cite{Wade2011}), the magnetic
field of HD\,191612 is indeed rather strongly positive in the phase 0.43. Clearly, more magnetic field observations 
of this star are needed 
to properly characterise the variation of the longitudinal magnetic field and the magnetic field geometry. 
In Fig.~\ref{fig:ofp191} we present our 
SOFIN observations together
with those published by Wade et al.\ (\cite{Wade2011}).

The supergiant HD\,151804 was previously observed three times in 2005 with FORS\,1 (Hubrig et al.\ \cite{Hubrig2008}).
No field detection at a significance level of 3$\sigma$ was achieved indicating that the field should be 
rather weak if present at all.
Our measurement  $\langle B_{\rm z}\rangle=-104\pm38$\,G based on HARPSpol observations confirms these 
previous results.

\section{Results for individual targets with detected magnetic fields}
\label{sect:indivi}

\subsection{HD\,36879}

\begin{figure}
\centering
\includegraphics[width=0.24\textwidth]{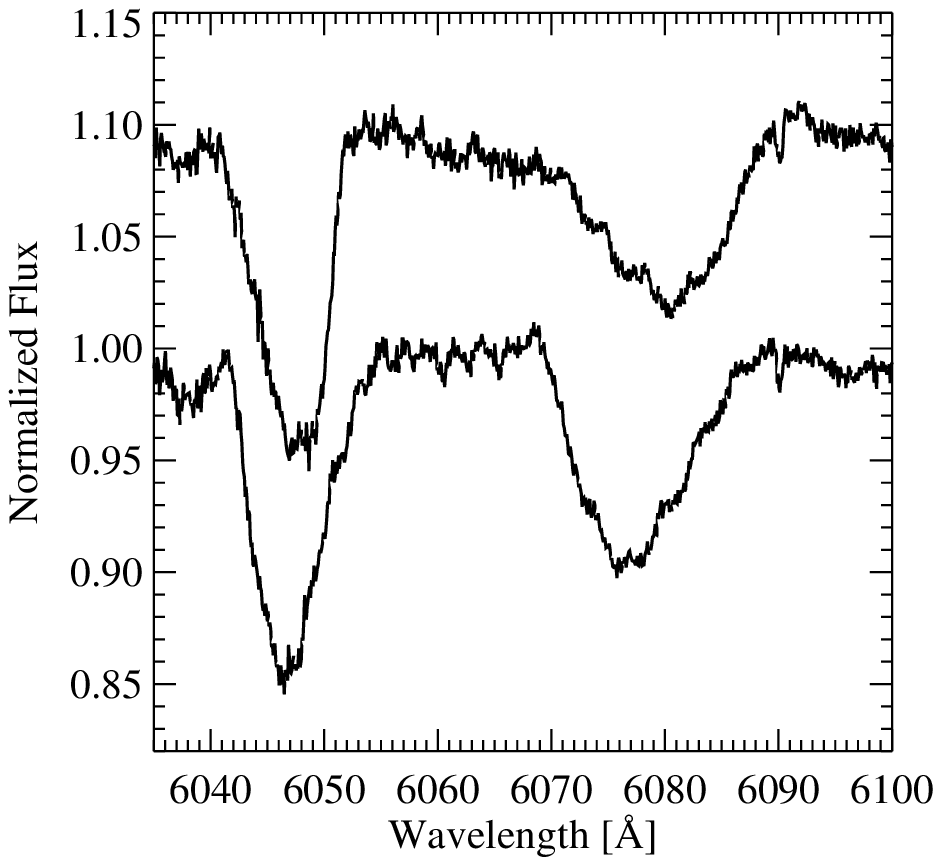}
\includegraphics[width=0.24\textwidth]{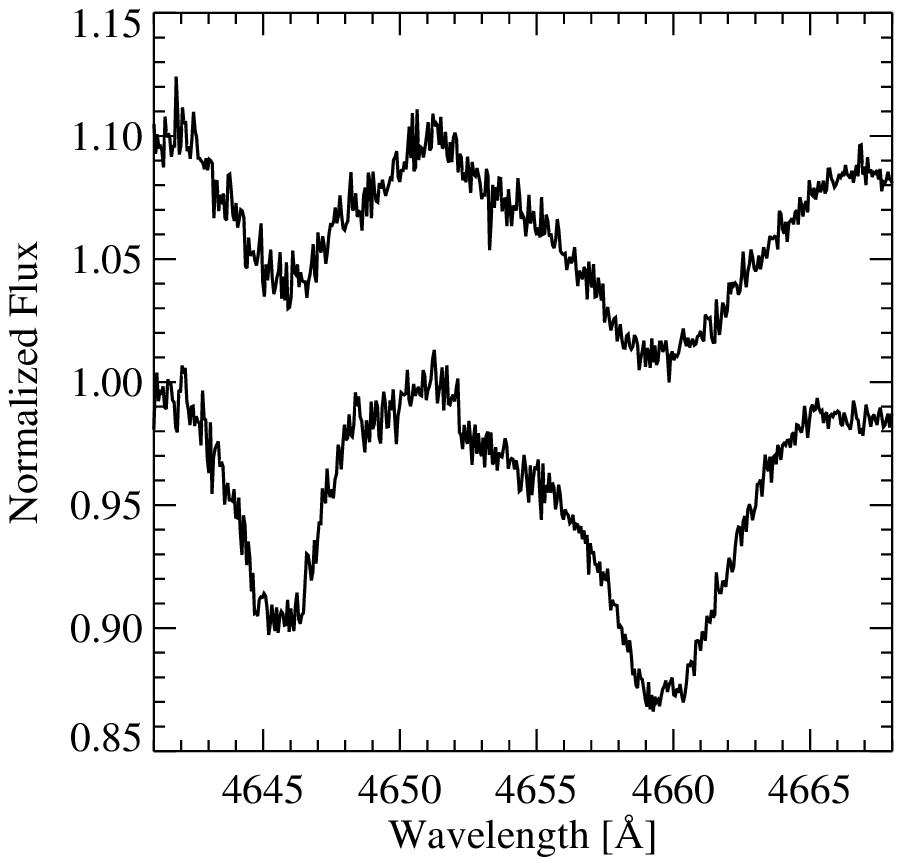}
\caption{Variable line profiles in the SOFIN spectra of HD\,36879 obtained at two different epochs.
{\sl Left panel:} Spectral variability of the \ion{Fe}{iii} line profile at $\lambda$6048 and the blend \ion{Fe}{iii}$+$\ion{He}{ii} at $\lambda$6078.
{\sl Right panel:} Spectral variability of line profiles presenting blends of the \ion{C}{iii} and \ion{O}{ii} lines at $\lambda$4647 and 
$\lambda$4660.
}
\label{fig:fe}
\end{figure}

\begin{figure}
\centering
\includegraphics[width=0.24\textwidth]{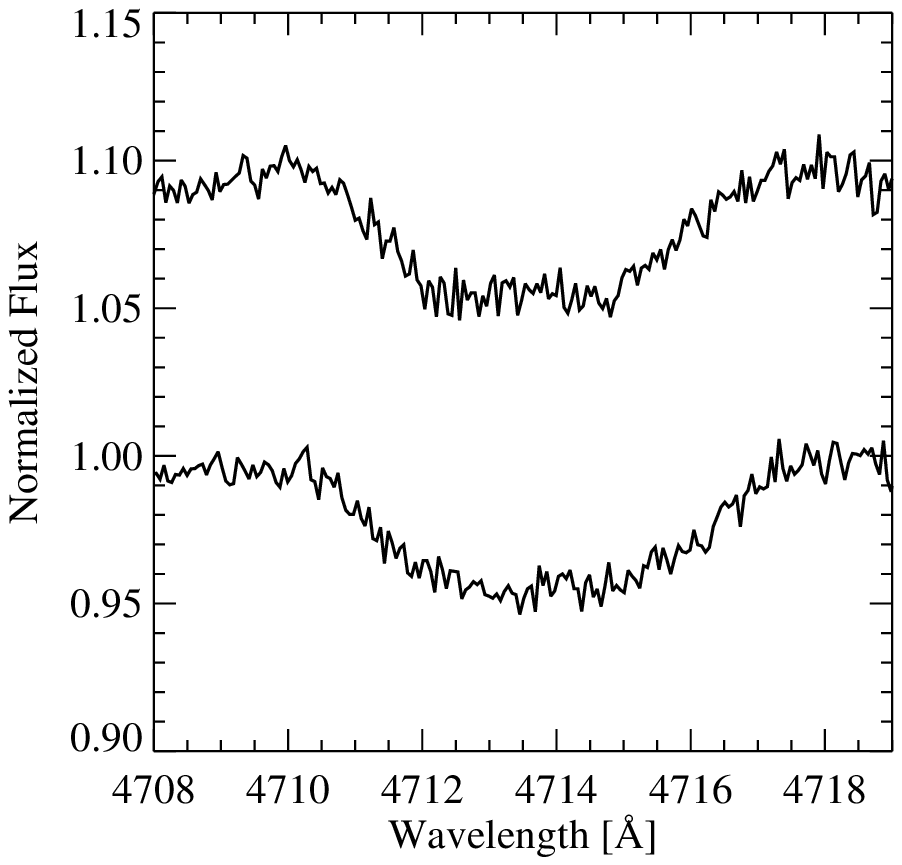}
\includegraphics[width=0.24\textwidth]{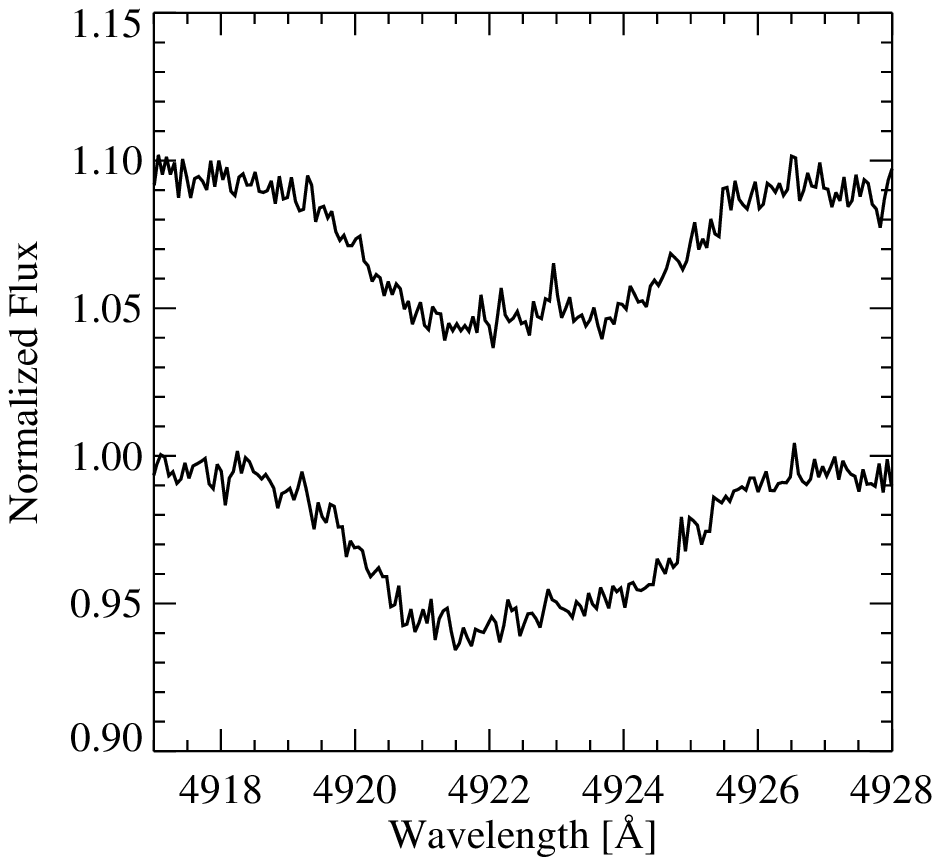}
\includegraphics[width=0.24\textwidth]{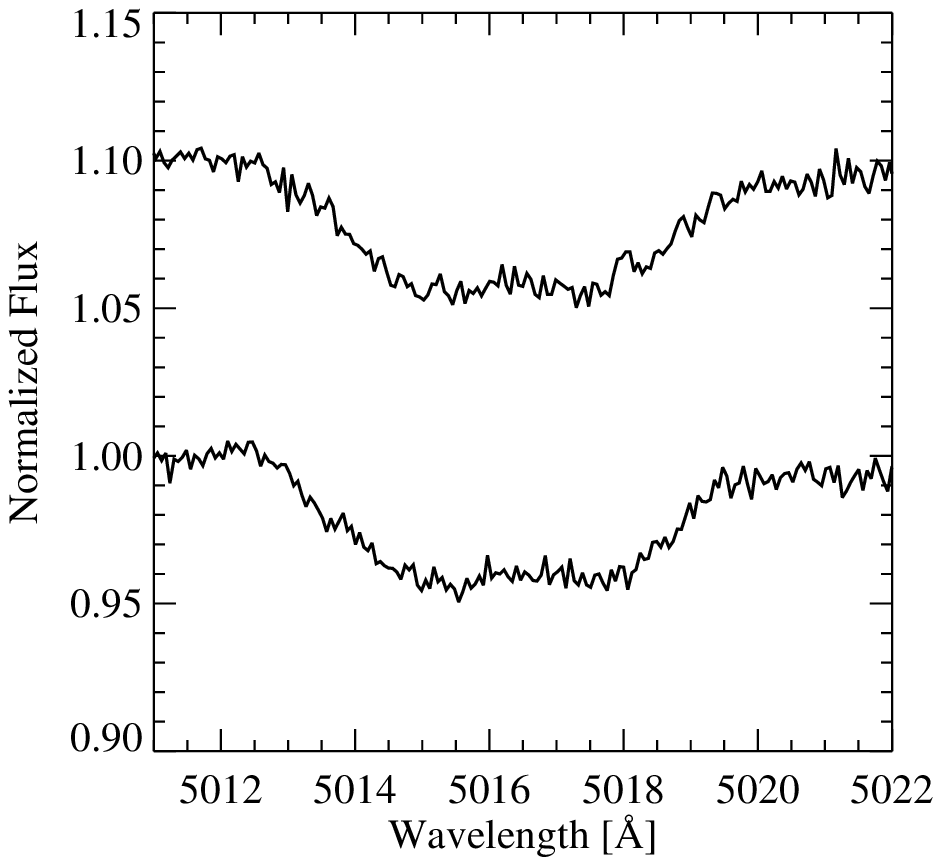}
\includegraphics[width=0.24\textwidth]{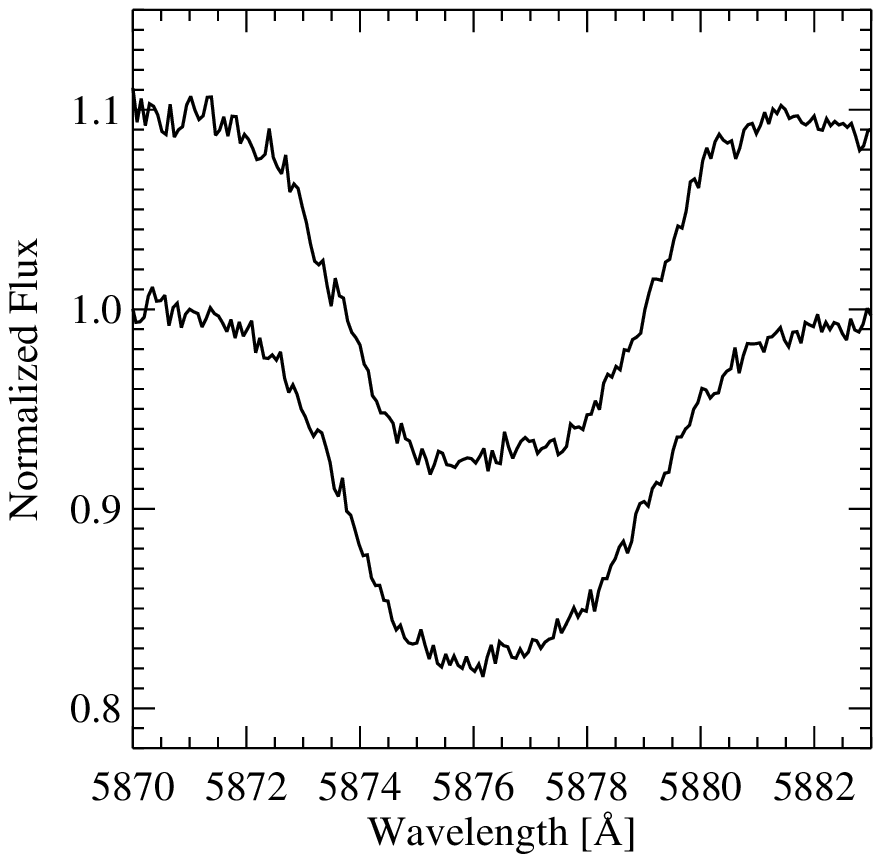}
\caption{Line profiles of \ion{He}{i} in the spectra of HD\,36879 obtained at two different epochs.
From top left to bottom right we show the lines \ion{He}{i} $\lambda$4713, $\lambda$4922, $\lambda$5015, and $\lambda$5876.
}
\label{fig:he}
\end{figure}

A magnetic field at a 3.5$\sigma$ significance level in HD\,36879 was for the first time detected in 2007
using FORS\,1 observations (Hubrig et al.\ \cite{Hubrig2008}).
Our most recent high-resolution SOFIN observations in 2009 December confirm the previous detection and 
reveal a change of the field polarity.
Walborn (\cite{Walborn2006}) reports that this star shows unexplained spectral peculiarities
and/or variations, among them peculiar narrow, variable \ion{Si}{iv} emission lines at $\lambda$1394 and $\lambda$1403
(e.g.\ Walborn \& Panek \ \cite{Walborn1984}). Distinct spectral variability is also 
discovered in \ion{Fe}{iii} and the blended lines \ion{C}{iii}/\ion{O}{ii} using our two high-resolution SOFIN spectra obtained on 
two different epochs.
The variations of \ion{Fe}{iii} and \ion{C}{iii}/\ion{O}{ii} line profiles are presented in Fig.~\ref{fig:fe}.
Interestingly, \ion{He}{i} lines appear slightly split. This appearance presented in Fig.~\ref{fig:he} could 
possibly indicate the presence of He spots on the 
stellar surface.  
%HD\,36879: He I symmetrical - double or emission? Also OIII 5592.
%All He I and O II (5592) lines are split on both phases indicating the presence of spots??.

\subsection{15\,Mon}

%The star 15\,Mon is classified as a pre-main sequence star in the SIMBAD database, probably due to its close 
%proximity to the Cone Nebula. 
Hubrig et al.\ (\cite{Hubrig2011b}) showed that this star is a probable member
of the young, rich  open cluster NGC\,2264 located in the Monoceros OB\,1 association. 
From the spatial distribution of H$\alpha$ emission stars, Sung et al.\ (\cite{Sung2008})
have concluded that the young open cluster
NGC\,2264 can be subdivided into three star forming regions (SFRs): two active
SFRs - the region around the bright emission nebula in
the southwest of the brightest star S\,Mon and the
region to the north of the famous cone nebula - and surrounding these an 
elliptical-shaped halo region.
The study of Teixeira et al.\ (\cite{Teixeira2012}) indicates that NGC\,2264 has undergone more 
than one star-forming event, where the
unembedded, insubstantial inner disk population and embedded thick inner disk population 
appear to have formed in separate episodes.
Markova et al.\ (\cite{Markova2004}) consider  15\,Mon as a Galactic O-type star 
with a mass of 32\,$M_{\odot}$ and $T_{\rm eff}=37\,500$\,K. Speckle measurements of 15\,Mon indicated the
presence of a companion of spectral type O9.5~V~n (e.g.\ Gies et al.\ \cite{Gies1993}). In very high S/N spectra becomes 
the companion noticeable by very broad underlying components to the strong \ion{He}{i} lines. 
According to Kaper et al.\ (\cite{Kaper1996}) and Walborn 
(\cite{Walborn2006}) shows this star
distinct peculiarities in the spectra, which could be typical for stars possessing magnetic fields. 
The X-ray luminosity based on {\em Chandra} observations is  $\log(L_{\rm X})=31.82$
and the ratio of X-ray to bolometric luminosity $\log(L_{\rm X}/L_{\rm bol}) = -6.63$
is slightly above the average  (Ramirez et al.\ \cite{Ramirez2004}, Naz\'e \cite{Naze2009}), 
as may be expected from a magnetic star. The high-resolution X-ray spectrum
of 15 Mon is typical for a star with its spectral type and, apparently, not
especially hard (Walborn et al.\ \cite{Walborn2009}). 

The previously 
measured magnetic field using the full FORS\,1 spectrum
coverage was of positive polarity at a 2.6$\sigma $ significance level (Hubrig et al.\ \cite{Hubrig2009a}).
A weak magnetic field was detected in this star at a significance level of 4.4$\sigma$ using SOFIN observations.
No contribution of the secondary component was detected in our FORS\,1 and SOFIN spectra (see also Fig.~\ref {fig:norm}). 

%15\,Mon: He I , N III, symmetrical, 

\subsection{CPD$-$28\,2561}

\begin{figure}
\centering
\includegraphics[angle=0,width=0.50\textwidth]{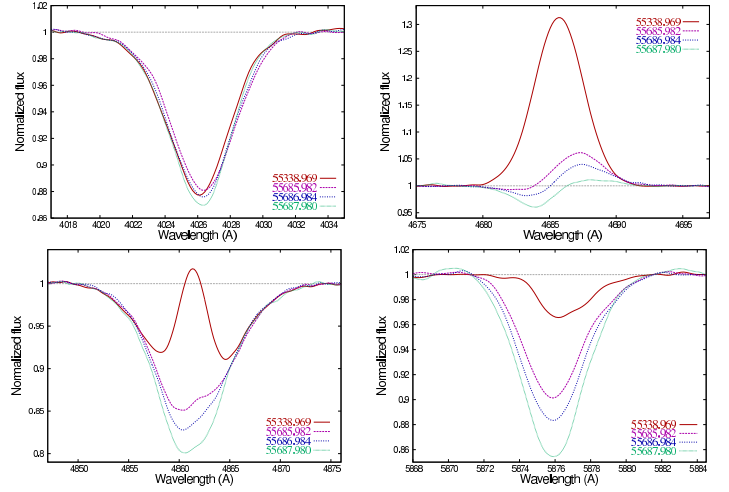}
\caption{
Distinct variability observed in the spectra of CPD$-$28\,2561 in the \ion{He}{i} $\lambda$4026 (top left),  
\ion{He}{ii} $\lambda$4686 (top right), H$\beta$ (bottom left), and  \ion{He}{i} $\lambda$5876 (bottom right) lines.
}
\label{fig:ofpCD}
\end{figure}

Our previous observations of the Of?p star CPD$-$28\,2561 enabled us to detect a magnetic field at the 3.1$\sigma$ level
using the whole spectrum and at the 3.2$\sigma$ level using Balmer lines (Hubrig et al.\ \cite{Hubrig2011b}).
Our new spectropolarimetric observations of this star, recently reported by us in a short IBVS article
(Hubrig et al.\ \cite{Hubrig2012a}) indicate that the magnetic field is variable. 
Due to the relative faintness of CPD$-$28\,2561 with $V$=10.1, it was only scarcely studied in the past. 
Levato et al.\ (\cite{Levato1988}) acquired radial velocities of 35 OB stars with carbon, nitrogen, and oxygen 
anomalies and found variability of a few emission lines with a probable period of 17\,days. Walborn et al.\ 
(\cite{Walborn2010}) mentioned that CPD$-$28\,2561 undergoes extreme spectral transformations very similar 
to those of HD\,191612, on a timescale of weeks, inferred from the variable emission intensity of the
\ion{C}{iii} $\lambda\lambda$~4647-4650-4652 triplet.
Our spectropolarimetric observations on three consecutive nights in 2011 reveal strong variations 
in several hydrogen and helium line profiles. 
The variation of \ion{He}{i} $\lambda$4026, \ion{He}{i} $\lambda$5876, and of 
H$\beta$ on such a short time scale were not reported in the literature before.
A few examples of the detected spectral line profile variations 
are presented in Fig.~\ref{fig:ofpCD}.
The contribution of a variable emission component is well visible in three lines.
The  profile of the \ion{He}{i} $\lambda$4026 line appears only weakly variable.
These variations 
have an amplitude of about 1--2\% of the continuum level and can likely be explained by 
rotation modulation.
The line \ion{He}{ii} $\lambda$4686 exhibited strong emission during our observations in 2010,
which transforms to a much weaker P\,Cygni line profile with a weak blue shifted absorption in 2011.
A contribution of the emission component is clearly seen in the profile of the \ion{He}{i} $\lambda$5876
line in 2010, but its presence can only be suspected in the variable line depth observed
in 2011.
A strong emission component is well visible in the line center of H$\beta$ in 2010. 
The shape of the line profile of H$\beta$ is similar to the behaviour of the 
H$\alpha$ line profile in the spectra
of 19\,Cep presented in Fig.~13 of Kaper et al.\ (\cite{Kaper1997}) and in Fig.~1
of Kholtygin et al.\ (\cite{Kholtygin2003}).

CPD$-$28\,2561 is an intrinsically bright X-ray source as detected by the {\em Rosat} 
X-ray observatory. Its ratio of X-ray to bolometric luminosity
($\log{L_{\rm X}/L_{\rm bol}}\approx -6$) makes it unusually X-ray bright
compared to other O-type stars (Chlebowski \cite{Chleb1989}, Oskinova \cite{Oskinova2005}). The hardness ratio
(${\rm HR1}\approx 0.04\pm 0.4$) indicates that its X-ray emission can be
harder than is typical in single O-type stars. This provides indirect support
for the presence of a magnetic field on the star, strong enough to confine
the stellar wind (Babel \& Montmerle \cite{Babel1997}).

\subsection{CPD$-$47\,2963}

For the star CPD$-$47\,2963, we confirm our previous 3.1$\sigma$ detection by measuring all absorption lines in 
the FORS\,2 spectra.
The magnetic field of this star is variable and shows a change of polarity over our four-night observing run.
According to Walborn et al.\ (\cite{Walborn2010}), CPD$-$47\,2963 belongs to the Ofc category, which consists of 
normal spectra 
with \ion{C}{iii} $\lambda\lambda$~4647-4650-4652 emission lines of comparable intensity to those of the Of defining lines 
\ion{N}{iii} $\lambda\lambda$~4634-4640-4642. 
The origin of the magnetic field in this star probably differs from those of other magnetic
O-type stars, because non-thermal radio emission, which is frequently observed
in binary systems with colliding winds, was detected by Benaglia et al.\ (\cite{Benaglia2001}). 
On the other hand, the membership of
CPD$-$47\,2963 to a binary or multiple system has not yet been investigated. 

\subsection{HD\,92206c}

The presence of a variable magnetic field up to a 4.4$\sigma$ significance level in HD\,92206c is documented 
here for the first time.
This star is a probable  member of the open cluster NGC\,3324 located inside of a partial ring of
nebulosity northwest of the Carina Nebula. The assessment of the membership using  astrometric catalogues with 
the highest quality kinematic and photometric data currently available is 
described in Sect.~\ref{sect:discussion}. 
According to the Galactic O Star Catalogue 
(Ma{\'{\i}}z-Apell{\'a}niz et al.\ \cite{MaizApellaniz2004}) belongs HD\,92206c to the triple system HD\,92206abc.
An analysis of the distribution of the ionised and neutral material associated with the \ion{H}{ii} region Gum~31
by Cappa et al.\ (\cite{Cappa2008}) suggested that the excitation sources of this region are  
the brightest stars in the open cluster NGC\,3324 with the triple system HD\,92206abc considered as the brightest 
cluster member. The detection of protostellar candidates almost coincident in position with the open cluster NGC\,3324
indicates that stellar formation is ongoing in this region. 
%which is the visual double star IDS 10336-5806 in the Index catalogue (Jeffers et al.\ 1963) with a 1 mag fainter 
%companion (HD\,92206B) placed 5 $^{\prime\prime}$ to the East. Another bright cluster member is located 
%35 $^{\prime\prime}$ to the SW. This star is CD-57$^\circ$3378, also referred to as HD\,92206C in the literature. 

Walborn (\cite{Walborn1982}) reports that the spectrum of HD\,92206c displays very strong, broad hydrogen lines,
possibly similar to those in the Orion Trapezium cluster, in particular $\theta^1$\,Ori\,C, and the appearance
of these lines is indicative of extreme youth. A search for a magnetic field was carried out in 
another member of this cluster, the bright A0-type supergiant 
HD\,92207. According to Hubrig et al.\ (\cite{Hubrig2012b}),
this star also possesses a magnetic field.

\subsection{HD\,93521}

A longitudinal magnetic field of the order of 200--300\,G is detected in this star at a significance level of 3.2$\sigma$.
This high Galactic latitude O9.5\,Vp star with $v$\,sin\,$i=390$\,km\,s$^{-1}$ (Rauw et al.\ \cite{Rauw2008}) is one 
of the fastest rotators in 
our sample, along with $\zeta$\,Oph. It is located at an unusually
high Galactic latitude of 62.15 and displays prominent line profile
variability in the optical and UV domain (Rauw et al.\ \cite{Rauw2008}). In the most recent work by Rauw et al.\ 
(\cite{Rauw2012a}), the authors  report that He and N are found to be overabundant and suggest that despite some ambiguities on the 
runaway status of the star it is likely that the star is a Population~I
massive O-type star that was ejected from the Galactic plane either through dynamical interactions or as the result of a 
supernova event in a binary system. 

HD\,93521 is one of the few O-type stars known to exhibit stellar pulsations. In this star, 
two frequencies of about 2--3\,hours are detected and interpreted as due to high-degree non-radial pulsation modes 
(Howarth \& Reid\ \cite{howarth_reid}; Rauw et al.\ \cite{Rauw2008}). 

The X-ray observations of this star have been summarized by Rauw et al.\ (\cite{Rauw2012a}). They concluded 
that the stellar X-ray spectrum is consistent with a normal 
late O-type star although with subsolar metallicity. No trace of a compact 
companion was found in the X-ray data. The ratio of X-ray to bolometric luminosity is
$\log(L_{\rm X}/L_{\rm bol}) = -7.1...-7.0$ and the stellar X-ray luminosity
$L_{\rm X}=5\times 10^{29}$\,erg\,s$^{-1}$ is rather low. Such low X-ray luminosities 
are also observed in other magnetic stars. 
Oskinova et al.\ (\cite{Oskinova2011}) showed, that relatively low X-ray luminosities can be 
explained in the framework of the magnetically confined wind shock model if the stellar
wind strength is correctly derived from an optical and UV spectroscopic analysis.  
Significantly, Rauw et al.\ (\cite{Rauw2012a}) show that besides a softer plasma component with
$kT_1=0.3$\,keV, which is typical for normal O-stars, a harder component with
$kT_2=3$\,keV is present in the X-ray spectrum of HD\,93521. Recently, Ignace et al.
(\cite{Ignace2012}) showed that such a hard component may serve as an indicator for the presence of a stellar magnetic 
field.

\subsection{HD\,93632 and HD\,93843}

A weak negative magnetic field in HD\,93632, just at a significance level of 3$\sigma$  was detected  
in one out of two FORS\,2 observations.
Walborn (\cite{Walborn1973}) reported that both HD\,93632 and HD\,93843 appear slightly variable in their spectral types.
Interestingly, also HD\,93843, similar to HD\,93632, exhibits a weak magnetic field. It's presence
is confirmed in our new observations.
According to Walborn et al.\ (\cite{Walborn2010}), the star HD\,93843 
belongs to the Ofc category.
Prinja et al.\ (\cite{Prinja1998}) monitored the stellar wind of this star using IUE time series.
They identified systematic changes in the absorption troughs of the \ion{Si}{iv} and \ion{N}{v} resonance 
lines with a repeatability 
of wind structures with a period of 7.1\,days.
The authors suggested at that time that a magnetic field might help to explain the cyclical wind perturbation.
Both stars have been observed during the {\em Chandra} X-ray survey of the Carina Nebula
(Townsley et al.\ \cite{Townsley2011}). For HD\,93632 is the X-ray luminosity $\log{L_{\rm X}=31.7}$.
The ratio of X-ray to bolometric luminosity is small, only 
$\log(L_{\rm X}/L_{\rm bol}) = -7.7$. A one-temperature fit to the X-ray 
spectrum shows a slightly higher than average temperature $kT=0.56$ (Naz\'e et al.\ \cite{Naze2011}).
For HD\,93843 is the X-ray luminosity $\log{L_{\rm X}=32.1}$.
The ratio of X-ray to bolometric luminosity is standard, 
$\log(L_{\rm X}/L_{\rm bol}) = -7.1$. A one-temperature fit to the X-ray 
spectrum derives a spectral temperature $kT=0.3$, which is lower than in 
HD\,93632 (Naz\'e et al.\ \cite{Naze2011}).  

%On the other hand, three other stars of the  Ofc category included in our survey,
%HD\,93204, HDE\,303308, and HD\,93403, do not have a detectable magnetic field at the 3$\sigma$ level.

\subsection{HD\,148937}

\begin{figure}
\centering
\includegraphics[width=0.45\textwidth]{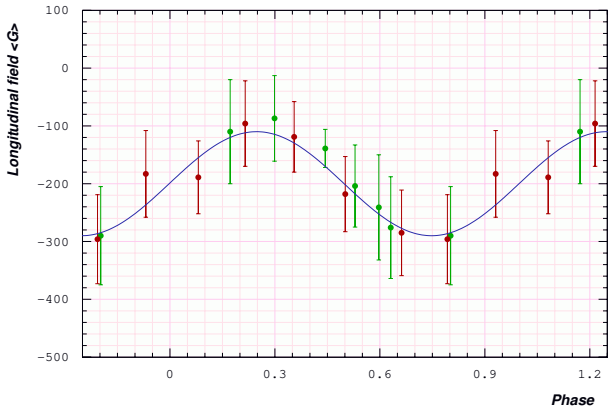}
\caption{
 Longitudinal magnetic field variation of the Of?p star HD\,148937 according to the 7.032\,d period 
determined by Naz\'e et al.\ (\cite{Naze2010}).
Red symbols correspond to the ESPaDOnS observations (Wade et al.\ \cite{Wade2012b}),
while green symbols are our FORS\,1 and FORS\,2 measurements
(Hubrig et al.\ \cite{Hubrig2008,Hubrig2011b} and present paper).
Note that the measurement errors for both ESPaDOnS and FORS\,1/2 observations are of similar order.
One more measurement (not shown in this Figure) was 
obtained in 2010 on May~22, but ordinary and extraordinary beams in the FORS\,2 setup were overlapping in this exposure 
due to problems with slitlets.
}
\label{fig:ofp}
\end{figure}

The first detection of a mean longitudinal magnetic field $\langle B_{\rm z}\rangle=-254\pm81$\,G
in the Of?p star HD\,148937 using FORS\,1 at the VLT was reported by Hubrig et al.\ (\cite{Hubrig2008}).
An extensive multiwavelength study of HD\,148937 was carried out by Naz\'e et al.\ (\cite{Naze2008}), who
detected small-scale variations of \ion{He}{ii} $\lambda$4686 and the Balmer lines with a period of
seven days and an overabundance of nitrogen by a factor of four compared to the Sun.
The periodicity of spectral variations in hydrogen and helium lines
was re-confirmed using additional higher resolution spectroscopic material indicating the
similarity to the other Of?p stars
HD\,108 and HD\,191612 (Naz\'e et al.\ \cite{Naze2010}).
Our new and old observations of HD\,148937 are presented in Fig.~\ref{fig:ofp} together with the ESPaDOnS 
observations obtained at the CFHT by Wade et al.\ (\cite{Wade2012b}).
This figure demonstrates the excellent potential of FORS\,2 for the
detection and investigation of magnetic fields in massive stars: 
While an exposure time of 21.5\,h at the CFHT was necessary to obtain seven binned measurements, the exposure 
time for FORS\,2 observations accounted only for two to four 
minutes and only 2.3\,h  were used 
for our observations at six different epochs including telescope presets and the usual overheads for readout time and 
retarder waveplate rotation.

\subsection{$\zeta$\,Oph}

\begin{figure}
\centering
\includegraphics[width=0.45\textwidth]{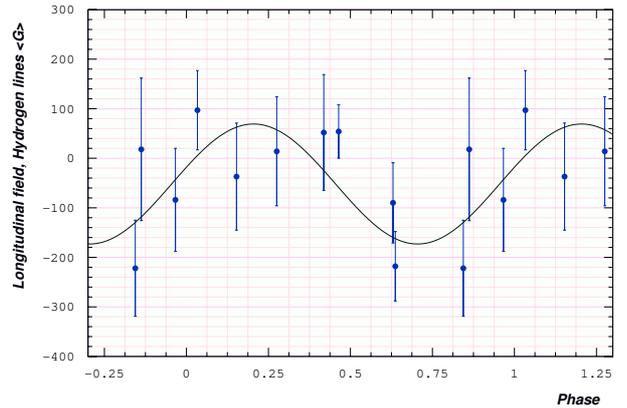}
\caption{
Phase diagram for the best sinusoidal fit corresponding to the period of 0.8\,days for the longitudinal magnetic field 
measurements using hydrogen lines for $\zeta$\,Oph.
}
\label{fig:oph_hydr08}
\end{figure}

\begin{figure}
\centering
\includegraphics[width=0.45\textwidth]{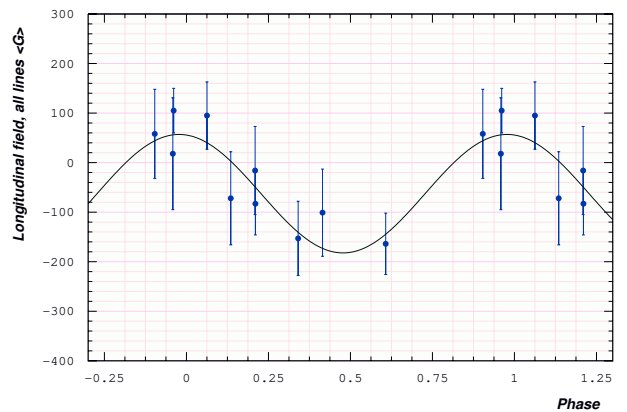}
\caption{
Phase diagram for the best sinusoidal fit corresponding to the period of 1.3\,days for the longitudinal magnetic field 
measurements using the whole spectrum for $\zeta$\,Oph.
}
\label{fig:oph_all1_32}
\end{figure}

The star $\zeta$\,Ophiuchi is a well-known rapidly rotating runaway star with extremely interesting 
characteristics. This star exhibits a wonderful bow shock as evident from IR {\em Spitzer} IRAC observations and 
undergoes episodic 
mass loss seen as emission in H$\alpha$. It is possible that $\zeta$\,Oph 
rotates with almost break-up velocity with $v$\,sin\,$i=400$\,km\,s$^{-1}$ (Kambe  et al.\
\cite{Kambe1993}). 
$\zeta$~Oph is the second target of our sample with detected non-radial pulsations with periods of hours, 
first discovered from ground-based observations (e.g.\ Kambe et al.\ \cite{kambe}). Later, the star was monitored 
by the {\em MOST} satellite, which revealed at least a dozen significant 
pulsation frequencies between 1 and 10\,d$^{-1}$, typical of $\beta$\,Cep-like pulsation modes 
(Walker et al.\ \cite{walker}). 

The first detection of a mean longitudinal magnetic field in this star
was announced by Hubrig et al.\ (\cite{Hubrig2011c}). Nine
additional FORS\,2 spectropolarimetric observations showing a change of polarity have been obtained 
over all four observing nights.
The rotation period of this star is still uncertain: Oskinova et al.\ (\cite{Oskinova2001}) studied
{\em ASCA} observations that covered just the expected rotation period of the
star of about 0.8--1.0\,d. A X-ray flux variability with $\sim$20\%\
amplitude  in the {\em ASCA} passband (0.5-10\,keV) with a period of 0.77\,d was reported. 
The X-ray spectra of $\zeta$\,Oph are not
especially hard as found from {\em Chandra} HETGS and {\em Suzaku} data
(Waldron \& Cassinelli \cite{Waldron2007}, Walborn et al.\ \cite{Walborn2009}).
%On the other hand, Waldron et al.\ (in preparation, private communication) found that
%{\em SUZAKU} data on $\zeta$\,Oph suggest a period of
%$\sim$0.98\,d that is consistent but slightly larger than
%the X-ray periodicities found in {\em ASCA} data (Oskinova et al.\ \cite{Oskinova2011})
%and in {\em Chandra} HETGS data (Waldron \cite{Waldron2005}).  In addition,
%the HETGS data appear to indicate an additional cyclic period of
%$\sim$8\,hours in the hard X-ray band ($>$1.2\,keV).  
The variation period is also not well-defined in our observations.  The resulting periodograms for the magnetic field
measurements using hydrogen lines shows a dominating peak corresponding to a period of about 0.8\,d, while
the measurements using the whole spectrum indicate a rotation period of about 1.3\,d.
In Fig.~\ref{fig:oph_hydr08} we present the 
magnetic field variation using hydrogen lines phased with the period of 0.8\,days, while  
Fig.~\ref{fig:oph_all1_32} refers to magnetic field variations determined from all lines phased with 1.3\,days.
Interestingly, just a few months ago 
Pollmann (\cite{Pollmann2012}) presented equivalent width changes of \ion{He}{i} $\lambda$6678 with a period
of 0.643\,d, which is roughly half of the magnetic period of 1.3\,d determined using the whole spectral region.
For the rotation period of $\sim$1.3\,d the equivalent width of the \ion{He}{i} line would display a double wave.
Such a behaviour of \ion{He}{i} lines is  frequently found in He variable early-type 
Bp stars (e.g.\ Briquet et al.\ \cite{Briquet2004}). 
Clearly, more measurements are needed to determine the true periodicity
of the magnetic field in this star. 
%The slope of continuum was removed - this is the reason why the old
%measurement has a different value now.

\begin{figure}
\centering
\includegraphics[width=0.40\textwidth]{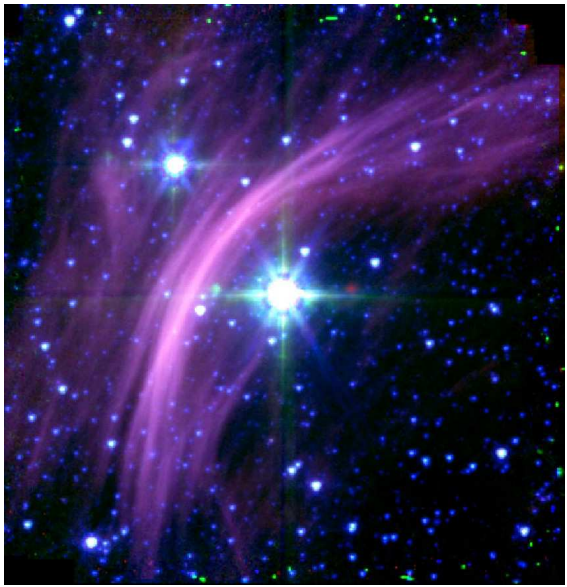}
\caption{Combined IR {\em Spitzer} IRAC (3.6\,$\mu$m blue, 4.5\,$\mu$m green, 8.0\,$\mu$m red) image of the bow shock 
around Vela\,X-1. Archival data have been used (AOR 17858304, PI: R. Iping). Image size is
$5' \times 5'$. 
}
\label{fig:bow}
\end{figure}

A bow-shock nebula is also present close to the high mass X-ray binary Vela\,X-1, consisting of a 
pulsar in an eccentric ($e = 0.09, P_{\rm orb} = 8.96$\,d) orbit around the here studied
B0.5 supergiant star HD\,77581.
In Fig.~\ref{fig:bow} we present an image of the bow-shock nebula based on archival {\em Spitzer}
IRAC maps (AOR 17858304, PI R.~Iping). The mass-loss rate ($\dot{M}\approx 10^{-6}\,M_\odot$\,yr$^{-1}$) for Vela\,X-1 is well 
constrained from its accretion powered X-ray luminosity (e.g., F\"urst et al.\ \cite{Furst2010}, Oskinova et al.\ 
\cite{Oskinova2012}).
Contrary to our study of $\zeta$\,Oph, no magnetic field detection at a significance level of 3$\sigma$ 
has been achieved in Vela\,X-1 during our observing run. 
However, we should keep in mind that the spectral behaviour of HD\,77581 is much more complex than that of 
$\zeta$\,Oph due to the presence of bumps and wiggles in the line profiles, and an impact of tidal effects producing 
orbital phase-dependent variations in the line profiles 
leading to asymmetries such as extended blue or red wings (Koenigsberger et al.\ \cite{Koenig2012}).
Furthermore, Kreykenbohm et al.\ (\cite{Krey2008}) detected flaring activity and temporary quasi-periodic 
oscillations in INTEGRAL X-ray observations.

\subsection{HD\,328856}

Our new observations confirm the presence of a magnetic field in this star.
As we already showed in our previous work on the basis of photometric membership probability, 
HD\,328856 is a member of the compact open
cluster Hogg\,22 in the Ara region at an age of 5\,Myr and a
distance of about 1300\,pc (Hubrig et al.\ \cite{Hubrig2011b}). On the other hand, its proper motions
indicate that HD\,328856 is not fully co-moving with the other cluster members, deviating from the cluster
mean proper motion by $\sim$2$\sigma$ (for more details of the membership probabilities,
see Kharchenko et al.\ \cite{Kharchenko2004}).

\subsection{CPD$-$46\,8221}

A search for the presence of a magnetic field in this star was carried out for the first time. A weak mean longitudinal
field was detected during the first observing night at a significance level of 3.1$\sigma$.
As we show below in Sect.~\ref{sect:discussion}, this star has a high kinematical membership probability in 
the open cluster Hogg\,22.

\subsection{HD\,157857}

Spectropolarimetric observations of this star have been conducted for the first time. Our measurements on the 
second observing night carried out on hydrogen lines resulted in a field detection at a significance level
of 3.7$\sigma$.
Schilbach \& Roeser (\cite{SchilbachRoeser2008}) retraced the orbit of HD\,157857 in the Galactic potential
and concluded that it was ejected from the open cluster NGC\,6611 about 3.8\,Myr ago with a peculiar velocity of 
113.5\,km\,s$^{-1}$.

\subsection{HD\,164794}

All four SOFIN and one HARPS high-resolution spectropolarimetric observations confirm our previous detection of 
the longitudinal magnetic field in this star. The field appears to be variable with negative and positive extreme values 
of $-$265\,G and $+$242\,G.
HD\,164794 belongs to the young cluster NGC\,6530 and according to the Galactic O Star Catalogue (Ma{\'{\i}}z-Apell{\'a}niz et al.\ 
\cite{MaizApellaniz2004}), has a classification O4~V~((f)), 
exhibiting weak \ion{N}{iii} emission and strong \ion{He}{ii} $\lambda$4686 absorption.
It is a well-known non-thermal radio 
emitter and, according to van Loo et al.\ (\cite{Loo2006}), the most likely mechanism is synchrotron emission from colliding 
winds, implying that all O stars with non-thermal radio emission should be members of binary or multiple systems. 
Hints of a wind-wind interaction were indeed detected in the X-ray domain (Rauw et al.\ \cite{Rauw2002}). 
A long-term study of its binary nature and spectrum variability has been recently presented by 
Rauw et al.\ (\cite{Rauw2012b}) who derived an orbital solution and an orbital period of 8.6\,yr.
%9\,Sgr: N is very week and variable (4510). too weak in the first two observations. He I 4713, 4921 (all nights)
%5015 almost absent, 
%lines are symmetrical. He I 4713, and N are stronger in the last two observations.

\subsection{HD\,191612}

Our SOFIN spectropolarimetric observations of this star have already been discussed in Sect.~\ref{subsect:sofin}.
HD\,191612 is currently one of the best studied targets in the optical and X-ray domains among
the five currently known Galactic Of?p stars. Due to the above mentioned discrepancies in the magnetic field
measurements around the positive field extremum, additional magnetic field observations
are necessary for proper magnetic field geometry modeling.
   
%HD\,191612 in the phase 0.43 He lines are very asymmetrical (core is shifted to the red),
%while in the phase 0.68 they are 
%almost symmetrical (the core is shifted to the blue). He I 4921 shows emission in the center in the 
%phase 0.43???
%NII lines are symmetrical in 
%phases 0.43 and 0.68. The \ion{C}{iv} lines ($\lambda$5801 and 5810) and the line 
%at $\lambda$5592.3 belonging to \ion{O}{iii} are symmetrical in all phases.

\section{Discussion}
\label{sect:discussion}

Whereas magnetic fields are ubiquitous in low mass stars due to their convective envelopes,
and found in 10...20\% of all intermediate mass stars, they were unexpected in O~stars until 
just a few years ago. 
The first O~star for which a magnetic field was detected is the spectroscopically variable
star $\theta^1$\,Ori~C (Donati et al.\ \cite{Donati2002}).
However, with ever more magnetic O~stars being identified in recent years 
(e.g., Hubrig et al.\ \cite{Hubrig2011b}, Martins et al.\ \cite{Martins2012}), it appears likely
that the as yet small number of magnetic O~stars is just a consequence of an observational bias,
due to the low number of spectral lines in these objects. Indeed, there may be more reasons to
find magnetic fields in O stars than in A~and B~stars. Fossil fields might appear with similar
frequencies on both groups. If close binary effects as mass transfer or stellar merger would 
produce large scale stable surface fields in O stars
(Ferrario et al.\ \cite{Ferrario2009}, Langer \cite{Langer2012}),
then about one quarter of them would be expected to be magnetic due to their very high close
binary fraction (Sana et al.\ \cite{Sana2012}). Furthermore, their proximity to the Eddington
limit produces convection in the envelopes of O~type stars, which could also produce 
surface magnetic fields (Cantiello et al.\ \cite{Cantiello2009}). 
%Only a few years ago, magnetic fields were unexpected in massive stars, due to the
%presumption of the absence of a sub-surface convective dynamo. 
%Quite recently, the presence of a convective zone in the outer envelopes of hot massive stars
%was studied by Cantiello et al.\ (\cite{Cantiello2009}) using a stellar evolution code to
%compute a grid of massive star models at different metallicities. They mapped the strength of the 
%iron convective zone (FeCZ) in the H-R diagram, showing its prominence as a function of stellar
%parameters. Since in their models FeCZ has a spatial extent similar to the solar convection zone,
%the authors suggest that a dynamo may also work in rapidly rotating OB stars.

With the present work we gradually increase the sample of massive stars with determined magnetic field strengths.
%%The accuracies of the field determinations using FORS\,1/2 and SOFIN are compared with those from 
%%ESPaDOnS observations for the stars with known field variation curves revealing that our measurement errors 
%%are in general of the same order.
It is the aim of our survey to provide new clues about the origin and detectability
of the magnetic fields of massive stars.
Bagnulo et al.\ (\cite{Bagnulo2012}) used the ESO FORS\,1 pipeline
to reduce the full content of the FORS\,1 archive.
They claim that very small instrument flexures, negligible in most
of the instrument applications, may be responsible for some spurious
magnetic field detections.
%As already stated in Sect.~\ref{sect:FORS_obs}, the strongest
%magnetic fields are detected in the two Of?p stars CPD$-$28\,2561 and  HD\,148937.
We note however that
out of the five Of?p stars, two were detected to be magnetic for the first time
by us (Hubrig et al.\ \cite{Hubrig2008,Hubrig2011b}).
The excellent agreement of our FORS\,1 and FORS\,2 data with ESPaDOnS 
observations by Wade et al.\ (\cite{Wade2012b}) of HD\,148937 can be seen in Fig.~\ref{fig:ofp}.
Bagnulo et al.\ (\cite{Bagnulo2012}) could not detect the magnetic field
in 9\,Sgr (=HD\,164794), while our high resolution observations of this star
with both SOFIN and HARPS fully confirm our earlier result (Hubrig et al.\ \cite{Hubrig2008}).
Also, high resolution observations with ESPaDOnS (Petit et al.\ \cite{Petit2013}) 
showed a weak magnetic field in 40\,Tau (=HD\,25558), which was already detected by
Hubrig et al.\ (\cite{Hubrig2009b}), but not found by Bagnulo et al.\ (\cite{Bagnulo2012}) 
from the same data.
Overall, it is not clear why Bagnulo et al.\ (\cite{Bagnulo2012}) are not able
to find magnetic fields found by other groups with FORS\,1/2 and other instruments.
These are only a few examples that
show the need for confirmations of FORS\,1 magnetic field measurements with
high spectral resolution polarimeters.
In the following, we discuss the role of
magnetic stars in star clusters, magnetic runaway stars, and indirect magnetic field indicators
in massive stars. 
%Our survey of magnetic fields in massive stars provides new clues about the origin of their magnetic fields.
%Magnetic fields are detected in stars of different classification
%categories, with different kinematical status, and at different evolutionary stages.
%Clearly, magnetic field detections in the five Of?p stars imply that there is a tight relation between the 
%observed properties of the Of?p star group and the presence of a magnetic field. 
%Apart from the Of?p star NGC\,1624-2, no other Of?p star belongs to an open cluster or association.
%The analysis of the stellar content of the cluster NGC\,1624 by Jose et al.\ (\cite{Jose2011}) indicated that
%the Of?p star NGC\,1624-2 is the main ionizing source of the region, and the maximum post-main-sequence age of 
%the cluster is estimated as $\sim$4\,Myr.

\subsection{Magnetic stars in young clusters}
\begin{table*}
\caption{
Probable members in open clusters. 
}
\label{tab:clusters}
\centering
%\begin{tabular}{lrrrlrrrrrrrr}
\begin{tabular}{lrrrlrrr@{,\,}r@{,\,}rr@{,\,}r@{,\,}r}
\hline
\hline
\multicolumn{1}{c}{Object} &
\multicolumn{1}{c}{ASCC} &
\multicolumn{1}{c}{$P_{\rm kin}$} &
\multicolumn{1}{c}{$P_{\rm phot}$} &
\multicolumn{1}{c}{Cluster} &
\multicolumn{1}{c}{dist} &
\multicolumn{1}{c}{log\,$t$}& 
\multicolumn{3}{c}{$(\mu_X$, $\mu_Y$, $\sigma_{\mu})^{\rm star}$} &
%\multicolumn{1}{c}{$PM_Y^{star}$}&
%\multicolumn{1}{c}{$\sigma_{PM}^{star}$}&
%\multicolumn{1}{c}{$PM_X^{cl}$}&
%\multicolumn{1}{c}{$PM_Y^{cl}$}&
\multicolumn{3}{c}{$(\mu_X$, $\mu_Y$, $\sigma_{\mu})^{\rm cl}$}\\
\multicolumn{1}{c}{name} &
\multicolumn{1}{c}{number} &
\multicolumn{1}{c}{[\%]} &
\multicolumn{1}{c}{[\%]} &
\multicolumn{1}{c}{ } &
\multicolumn{1}{c}{[pc]} &
\multicolumn{1}{c}{[yr]} &
%\multicolumn{1}{c}{[mas/y]} &
%$\multicolumn{1}{c}{[mas/y]} &
\multicolumn{3}{c}{[mas/yr]} &
%\multicolumn{1}{c}{[mas/y]} &
%\multicolumn{1}{c}{[mas/y]} &
\multicolumn{3}{c}{[mas/yr]} \\
\hline
$^{*}$HD\,47839 & 1021435 & 62 & 100 & NGC\,2264 &  660 & 6.81 & $-$3.84 & $-$2.50 & 0.94 & $-$2.70 & $-$3.50 & 0.25 \\
HD\,92206a & 2231158 & 80 &  96 & NGC\,3324 & 2301 & 6.72 &$-$10.32 &    6.76 & 4.05 & $-$7.81 &    3.77 & 0.64 \\
HD\,92206b & 2231162 &  0 &  12 & NGC\,3324 & 2301 & 6.72 &   49.61 &    0.86 & 3.70 & $-$7.81 &    3.77 & 0.64 \\
$^{*}$HD\,92206c & 2231147 & 90 &  97 & NGC\,3324 & 2301 & 6.72 &$-$8.87 &    4.57 & 1.90 & $-$7.81 &    3.77 & 0.64 \\
%{\it HD\,92207} & 2231181 & 87 & 100 & NGC\,3324 & 2301 & 6.72 &$-$ 7.18 &    3.43 & 7.28 & $-$7.81 &    3.77 & 0.64 \\
HD\,93026 & 2232306 & 86 &  88 & Bochum\,10 & 2027 & 6.93 & $-$3.74 &    3.69 & 4.23 & $-$6.61 &    2.56 & 0.62 \\
CPD$-$58\,2611 & 2232406 & 96 & 100 & Trumpler\,14 & 2753 & 6.67 & $-$3.09 &    4.44 & 2.66 & $-$3.91 &    3.65 & 0.60 \\
$^{*}$HD\,328856 & 2074567 & 22 & 100 & Hogg\,22 & 1297 & 6.70 &    0.78 & $-$1.37 & 1.26 & $-$0.84 & $-$4.39 & 0.43 \\
$^{*}$CPD$-$46\,8221 & 2074573 & 97 &   1 & Hogg\,22 & 1297 & 6.70 & $-$0.45 & $-$5.19 & 1.32 & $-$0.84 & $-$4.39 & 0.43 \\
HD\,150958 & 2074576 & 93 &  84 & Hogg\,22 & 1297 & 6.70 & $-$0.95 & $-$3.67 & 1.20 & $-$0.84 & $-$4.39 & 0.43 \\
HD\,153919 & 1877055 &  0 & 100 & NGC\,6281 &  494 & 8.51 &    1.42 &    4.84 & 1.04 & $-$2.96 & $-$3.75 & 0.23 \\
HD\,156041 & 1878768 & 98 &  63 & Bochum\,13 & 1077 & 7.08 &    0.03 & $-$0.74 & 1.70 & $-$0.28 & $-$1.20 & 0.30 \\
HD\,156154 & 1878846 & 82 & 100 & Bochum\,13 & 1077 & 7.08 & $-$0.88 & $-$2.53 & 1.64 & $-$0.28 & $-$1.20 & 0.30 \\
$^{*}$HD\,164794 & 1593528 & 69 & 100 & NGC\,6530 & 1322 & 6.67 &    1.92 & $-$0.40 & 1.09 &    2.01 & $-$1.81 & 0.51 \\  
  \hline 
 \hline
\end{tabular}
\tablefoot{
%\flushleft\textbf{Note.}
 For each star, we give in the first two columns the object name and the corresponding ASCC catalogue number.
The kinematic and photometric probabilities for cluster membership presented in Cols.~3 and 4 were calculated 
according to the procedures described by Kharchenko et al.\ (\cite{Kharchenko2004}). 
The cluster names, the distances, and
the ages are presented in the next three columns. The proper motions and their errors for stars and clusters 
are given in
the last columns. All cluster data are taken from Kharchenko et al.\ (\cite{Kharchenko2005}).
Stars with detected magnetic fields are marked with an asterisk in the first column.
}
\end{table*}

Of the 30 massive stars analysed in this paper, 
eleven stars have been found to be related to open clusters of different ages. 
The data on the cluster membership of
these probable cluster O-type stars are presented in Table~\ref{tab:clusters}.
As database for the compilation of Table~\ref{tab:clusters}, we used the All-sky Compiled Catalogue of 2.5 million 
stars (ASCC-2.5, 3rd version) of Kharchenko \& Roeser (\cite{KharchenkoRoeser2009}).
The kinematic and photometric probabilities for cluster membership presented in Cols.~3 and 4 were calculated 
according to the procedures described by Kharchenko et al.\ (\cite{Kharchenko2004}). 
For calculating the kinematic membership probability, only proper motion information was used. 
Since the star HD\,92206c in the open cluster NGC\,3324 belongs to the triple system HD\,92206abc, we present 
the proper motions of all triple system members in the same table. 
Largely deviating proper motions of the cluster member HD\,92206b 
in comparison to the average cluster proper motions are
very likely caused by the orbital motion in the pair HD\,92206ab. Interestingly, not only the cluster member
HD\,92206c shows the presence of a magnetic field. According to Hubrig et al.\ (\cite{Hubrig2012b}), also another 
cluster member, the supergiant HD\,92207, possesses a magnetic field of the order of a few hundred Gauss.
Clearly, the cluster NGC\,3324 should be included in future magnetic field surveys of massive stars to 
better understand the role of the environment on the generation of their magnetic fields.      

As we already mentioned in our previous work, according to Schilbach \& Roeser (\cite{SchilbachRoeser2008}),
the star HD\,153919 was probably ejected from the cluster NGC\,6231 at an age of about 6.5\,Myr.
At present, HD\,153919 is located 4.2$^\circ$ from the center of NGC\,6231.
Considering the proper motions of HD\,153919 given by Kharchenko \&
Roeser (\cite{KharchenkoRoeser2009};
$\mu_\alpha=+1.42\pm1.17$\,mas/yr, $\mu_\delta=+4.84\pm0.87$\,mas/yr) 
and those of the cluster by Kharchenko et al.\ (\cite{Kharchenko2005};
$-0.39\pm0.52$\,mas/yr, $-1.99\pm0.38$\,mas/yr), then 2.15\,Myr ago the star was at $0.55\pm0.95^\circ$ 
from the cluster center.

Among the stars presented in Table~\ref{tab:clusters}, only five stars show the presence of magnetic fields,
HD\,47839, HD\,92206c, HD\,328856, CPD$-$46\,8221, and HD\,164794. Two of them, HD\,328856 and CPD$-$46\,8221,
belong to the open cluster Hogg\,22, which appears as a second promising cluster for investigating the magnetic nature
of cluster members. Although we deal here with small number statistics, we cannot refrain from mentioning that
these five stars with detected magnetic fields belong to the youngest clusters in our small sample. 

\subsection{Magnetic massive runaway stars}

Magnetic fields are detected in stars of spectral types,
with different kinematical status, and at different evolutionary stages.
Clearly, the detection of magnetic fields in all five Galactic Of?p stars implies a tight relation between the
spectral characteristics of the Of?p star group and the presence of a magnetic field. Remarkably, 
apart from the Of?p star NGC\,1624-2, no other Of?p star belongs to an open cluster or association.
The analysis of the stellar content of the cluster NGC\,1624 by Jose et al.\ (\cite{Jose2011}) indicated that
NGC\,1624-2 is the main ionizing source of the region, and the maximum post-main-sequence age of
the cluster is estimated as $\sim$4\,Myr.

Furthermore, recent studies of the evolutionary state of several magnetic massive stars indicate that
some of them evolved significantly or have the status of a runaway star (e.g., Martins et al.\ \cite{Martins2010}, 
Hubrig et al.\ \cite{Hubrig2011b,Hubrig2011d}). 
In the literature, the runaway status or candidate runaway status 
(i.e.,\ if stars possess high space velocities determined through proper motion and/or radial velocity measurements)
is assigned to several
magnetic massive stars, such as $\zeta$\,Oph (Blaauw \cite{Blaauw1952}), HD\,93521 (Gies \cite{Gies1987}),
HD\,157857 (Schilbach \& Roeser \cite{SchilbachRoeser2008}), HD\,57682 (Comeron et al.\ \cite{Comeron1998}), and 
the three 
Of?p stars HD\,108, HD\,148937, HD\,191612 (Hubrig et al.\ \cite{Hubrig2011d}).
It is known that some of the massive stars located far from star clusters and star-forming regions are runaways. 
They were likely formed in embedded clusters and then ejected into the field because of dynamical few-body 
interactions or binary-supernova explosions (Eldridge et al.\ \cite{Eldridge2011}). 
Although the group of field O stars whose runaway status is difficult to 
prove does exist, the recent study of Gvaramadze et al.\ (\cite{Gvaram2012}) showed no significant evidence in support 
of the in-situ proposal for the origin of isolated massive stars.
%On the other hand, recent studies of the evolutionary state of several magnetic massive stars indicate that
%some of them evolved significantly or have runaway status (e.g., Martins et al.\ \cite{Martins2010}, 
%Hubrig et al.\ \cite{Hubrig2011a,Hubrig2011b}). 
%In the literature, the runaway status or candidate runaway status 
%(i.e.\ if stars possess high space velocities determined through proper motion and/or radial velocity measurements)
%is assigned to several
%magnetic massive stars, such as $\zeta$\,Oph (Blaauw \cite{Blaauw1952}), HD\,93521 (Gies \cite{Gies1987}),
%HD\,157857 (Schilbach \& Roeser \cite{SchilbachRoeser2008}), HD\,57682 (Comeron et al.\ \cite{Comeron1998}), and the three 
%Of?p stars HD\,108, HD\,148937, HD\,191612 (Hubrig et al.\ \cite{Hubrig2011c}).
%It is known that some of the massive stars located far from star clusters and star-forming regions are runaways. 
%They were likely formed in embedded clusters and then ejected into the field because of dynamical few-body 
%interactions or binary-supernova explosions. Although the group of field O stars whose runaway status is difficult to 
%prove does exist, the recent study of Gvaramadze et al.\ (\cite{Gvaram2012}) showed no significant evidence in support 
%of the in-situ proposal on the origin of massive stars.

In some cases, even if the star is considered as an open cluster member, a careful study of their
kinematic characteristics indicates that it is not fully co-moving with the other cluster members.
As an example, the space motion of the magnetic O-type star $\theta^1$\,Ori\,C 
was studied by van Altena et al.\ (\cite{vanAltena1988}), who reported that $\theta^1$\,Ori\,C is 
moving at 4.8$\pm$0.5\,km\,s$^{-1}$
towards position angle 142$^{\circ}$ and that this velocity is significantly larger than the dispersion 
value of 1.5$\pm$0.5\,km\,s$^{-1}$ found for 
the other cluster members. The results of the radial velocity study of Stahl et al.\ (\cite{Stahl2008})
indicate that this star is moving rapidly away from the Orion Molecular Cloud and its host cluster.
 Another example is the kinematical status of the magnetic O-type star HD\,328856, which is on the basis of the 
photometric membership probability a member of the compact open
cluster Hogg\,22 in the Ara region at an age of 5\,Myr and a
distance of about 1300\,pc. However, its proper motions
indicate that HD\,328856 is not fully co-moving with the other cluster members, deviating from the cluster
mean proper motion by $\sim$2$\sigma$ (Hubrig et al.\ \cite{Hubrig2011b}).

Among our studied runaway stars, HD\,93521 and $\zeta$\,Oph are magnetic fast rotators with $\beta$\,Cep-type 
pulsations. These stars are thus particularly interesting to calibrate stellar rotation models including magnetic 
fields in the hot upper part of the Hertzsprung-Russell diagram. Variability linked 
to stellar pulsation has never 
been searched 
for or found in any of our other sample targets. In fact, there are only a few O-type stars for which convincing 
evidence of non-radial pulsations with periods of a few hours has been detected from ground-based observations. 
Besides HD\,93521 and $\zeta$\,Oph, there are $\xi$\,Per and $\lambda$\,Cep (de Jong et al.\ \cite{dejong}). 
For such stars, the photometric amplitudes are below mmag level, which are very difficult to detect from the ground. 
It may explain why only few pulsating O-type stars have been identified so far and mostly spectroscopically. 
Recently, the very high-precision photometry of the {\em CoRoT} satellite 
led to the detection of stellar pulsation in three other O-type objects that belong to the young open cluster 
NGC\,2244 inside the Rosette nebula and to the surrounding association Mon\,OB2. $\beta$\,Cep-like pulsation modes 
have been found in HD\,46202 (Briquet et al.\ \cite{briquet11}) and in HD\,47129 (Mahy et al.\ \cite{mahy}) while 
unexpected modes of stochastic nature have been detected in HD\,46149 (Degroote et al.\ \cite{degroote}). Efforts 
to discover magnetic massive pulsators are worthwhile since their study provides a unique opportunity to probe the 
impact of a magnetic field on the physics of mixing inside stellar interiors.

\subsection{Indirect B-field indicators}

One of the major characteristics of magnetic Ap and Bp stars is the spectral variability due to the presence of 
chemical spots showing a certain symmetry with respect to the magnetic field geometry. Although spectral variability is
present in all discovered magnetic massive stars, 
the interpretation of line profile variations seems not to be 
as simple as for Ap and Bp stars: the behaviour of line profiles of different ions of a given element 
does not clearly correspond to the abundance variations. Synthetic spectra calculations of the line profile 
variations in massive stars are furthermore more troublesome since different spectral lines are to a different degree 
sensitive to 
the stellar wind and to non-LTE effects. In any case, a detection of line profile variations can certainly be used 
as an indirect indicator for the possible presence of a magnetic field. 
Another aspect that could hint at the presence of a magnetic field in massive stars 
is the fact that a number of magnetic OB-type stars show nitrogen 
enrichment (e.g.\ Morel et al.\ \cite{Morel2008}).

It is not clear yet how to use the X-ray emission as an indirect indicator for the 
presence of magnetic fields.
Anomalous X-ray emission can be expected from a massive star with a magnetic 
field strong enough to regulate the stellar wind motion.
The relative importance of magnetic fields in gases is described by the plasma-$\beta_{\rm p}$ parameter 
with $\beta_{\rm p}=8\pi p/B^2$, where $p$ is the gas pressure.  The gas is magnetically 
dominated when $\beta_{\rm p}<1$. For supersonic flows, such as stellar winds, the ram 
pressure exceeds the gas pressure and the dynamical importance of a magnetic field is 
defined by the ratio of wind kinetic to magnetic energy density: when this ratio is 
less than one, the magnetic field dominates the wind bulk motion. 

Oskinova et al.\ (\cite{Oskinova2011}) investigated the X-ray emission and wind properties among the 
complete sample of known magnetic early B-type stars with available X-ray observations. 
The authors have found that the X-ray emission from magnetic B-type stars is diverse. 
While some stars display hard variable X-rays, others are rather soft sources. The X-ray 
luminosities differ among otherwise similar Bp stars by more than two orders of magnitude.
Among magnetic O-type stars, the situation is similar. While some of them are bright, hard, 
variable X-ray emission as predicted by the magnetically confined wind-shock scenario
(e.g.\ Schulz et al.\ \cite{Schulz2000}), others are relatively soft and faint X-ray sources 
(Naz\'e et al.\ \cite{Naze2010}). 

Thus, the analysis of X-ray observations of magnetic OB stars shows that some of them 
are lacking strong and hard X-ray emission, or even are X-ray dark. Therefore, while a 
strong hard X-ray emission serves as a good indicator for the potential presence of 
a strong stellar magnetic field, X-ray strong stars represent only a sub-class of all 
magnetic massive stars. 

%X-ray observations for the magnetic stars presented in this paper also confirm 
%the lack of a clear trend between stellar magnetism and X-ray emission. New observation and modeling 
%efforts are required to fully understand the relations between magnetism and X-rays in massive stars.

\section{Outlook}

What is the origin of the magnetic fields in massive stars?
It is quite possible that massive O-type stars behave similar to slowly
rotating, chemically peculiar, magnetic Ap and Bp stars, which amount for 10 to 20\% of main-sequence
A and B stars. It was suggested that their magnetic fields 
are inherited from their parent molecular clouds (Moss \cite{Moss2003}).
Alternatively, from analysing possible paths for the formation of Ap and Bp stars
based on modern theories for the evolution of single and binary stars,
Tutukov \& Fedorova (\cite{Tut2010}) suggested that merging of close binaries
is their main formation channel.

Indeed, the evolution of a high fraction of massive stars
is affected by environmental effects. Sana et al.\ (\cite{Sana2012})
find two thirds of all Galactic O stars to be members of close binary systems.
Mass transfer or stellar merger
may rejuvenate the mass gaining star while strong differential rotation may cause
the creation of a magnetic field (Langer \cite{Langer2012}). In dense star clusters,
close multibody interactions may cause single and binary stars to be scattered
out of the star cluster (e.g.\ Leonard \& Duncan \cite{Leon1990}).

Clearly, these various ideas bear significantly different predictions for the appearance of magnetic main
sequence stars inside and outside of star clusters. This means that these predictions are well testable,
and we are confident that the question of the origin of the magnetic fields in massive
main sequence stars will be answered in the near future. 
As described in Sect.~\ref{sect:discussion}, first trends are emerging from the scarce data at hand.
However, at the current stage, we clearly need more observations before firm conclusions can be drawn.
 
{
\acknowledgements
This research made use of the SIMBAD database,
operated at CDS, Strasbourg, France.
The Nordic Optical Telescope is operated on the island of La Palma jointly by  Denmark, Finland, Iceland, 
Norway, and Sweden, in the Spanish Observatorio del Roque de los Muchachos of the Instituto de 
Astrofisica de Canarias.
We would like to thank G.~Lo~Curto for his help with the HARPS data reduction, and V.~Gvaramadze for 
stimulating discussions.
SH and JFG acknowledge support by the Deutsche Forschungsgemeinschaft (Hu532/17-1).
}


\begin{thebibliography}{}

\bibitem[1997]{Babel1997}
Babel, J., \& Montmerle, Th.\ 1997,
ApJ, 485, 29 

\bibitem[2012]{Bagnulo2012}	
Bagnulo, S., Landstreet, J.~D., Fossati, L., \& Kochukhov, O.\ 2012,
A\&A, 538, A129

\bibitem[2001]{Benaglia2001}
Benaglia, P., Cappa, C.~E., \& Koribalski, B.~S.\ 2001,
A\&A, 372, 952

\bibitem[1952]{Blaauw1952}
Blaauw, A.\ 1952,
Bull.\ Astron.\ Inst.\ Netherlands, 11, 414

%\bibitem[1961]{Blaauw1961}
%Blaauw, A.\ 1961,
%Bull.\ Astron.\ Inst.\ Netherlands, 15, 265

%\bibitem[2008]{BonannoUrpin2008}
%Bonanno, A., \& Urpin, V.\ 2008,
%MNRAS, 388, 1679

\bibitem[2004]{Briquet2004}
Briquet, M., Aerts, C., L{\"u}ftinger, T., et al.\ 2004,
A\&A, 413, 273

\bibitem[2011]{briquet11} 
Briquet, M., Aerts, C., Baglin, A., et al.\ 2011, A\&A, 527, 112

%\bibitem[1996]{Brown1996}
%Brown, G.~E., Weingartner, J.~C., \& Wijers, R.~A.~M.~J.\ 1996,
%ApJ, 463, 297

\bibitem[2011]{Brott2011}
Brott, I., Evans, C. J., Hunter, I., et al.\ 2011, 
A\&A, 530, A116

\bibitem[2009]{Cantiello2009}
Cantiello, M., Langer, N., Brott, I., et al.\ 2009,
A\&A, 499, 279

\bibitem[2008]{Cappa2008}
Cappa, C., Niemela, V.~S., Amor\'in, R., \& Vasquez, J.\ 2008,
A\&A, 477, 173

\bibitem[1989]{Chleb1989}
Chlebowski, T.\ 1989,
ApJ, 342, 1091

\bibitem[1998]{Comeron1998}
Comeron, F., Torra, J., \& Gomez, A.~E.\ 1998,
A\&A, 330, 975

%\bibitem[2005]{deWit2005}
%de Wit, W.~J., Testi, L., Palla, F., \& Zinnecker, H.\ 2005,
%A\&A, 437, 247

%\bibitem[2002]{Dias2002}
%Dias, W.~S., Alessi, B.~S., Moitinho, A., \& L{\'e}pine, J.~R.~D.\ 2002,
%A\&A, 389, 871

\bibitem[2010]{degroote} 
Degroote, P., Briquet, M., Auvergne, M., et al.\ 2010, A\&A, 519, 38

\bibitem[1999]{dejong} 
de Jong, J. A., Henrichs, H. F., Schrijvers, C., et al.\ 1999, A\&A, 345, 172

\bibitem[1997]{Donati1997}
%Donati J.-F., Semel M., Carter B. D., Rees D. E., Collier Cameron A., 1997,
Donati, J.-F., Semel, M., Carter, B.~D., et al.\ 1997,
MNRAS, 291, 658

\bibitem[2002]{Donati2002}
Donati, J.-F., Babel, J., Harries, T. J., Howarth, I. D., Petit, P., Semel, M.\ 2002,
MNRAS, 333, 55

\bibitem[2006]{Donati2006}
Donati, J.-F., Howarth, I.~D., Bouret, J.-C., et al.\ 2006,
\mnras, 365, L6

\bibitem[2011]{Eldridge2011}
Eldridge, J. J., Langer, N., Tout, C. A., et al.\ 2011,
MNRAS, 414, 3501

\bibitem[2009]{Ferrario2009}
Ferrario, L., Pringle, J.E., Tout, C.A., Wickramasinghe, D.T.\ 2009,
MNRAS, 400, L1 

\bibitem[2010]{Furst2010}
F\"urst, F., Kreykenbohm, I., Pottschmidt, K., et al.\ 2010,
A\&A, 519, A37

\bibitem[1987]{Gies1987}
Gies, D.~R. \ 1987,
ApJS, 64, 545

\bibitem[1993]{Gies1993}
Gies, D.~R., Mason, B.~D., Hartkopf, W.~I., et al.\ 1993,
AJ, 106, 2072

%\bibitem[2002]{Girardi2002}
%Girardi, L., Bertelli, G., Bressan, A., et al.\ 2002,
%A\&A, 391, 195

%\bibitem[2004]{GomezCox2004}
%G{\'o}mez, G.~C., \& Cox, D.~P.\ 2004,
%ApJ, 615, 744

\bibitem[2012]{Gvaram2012}
Gvaramadze, V.~V., Weidner, C., Kroupa, P., \& Pflamm-Altenburg, J.\ 2012,
MNRAS, 424, 3037

\bibitem[1993]{howarth_reid} 
Howarth, I.~D., \& Reid, A.~H.~N.\ 1993, A\&A, 279, 148

\bibitem[2007]{Howarth2007}
Howarth, I.~D., Walborn, N.~R., Lennon, D.~J., et al.\ 2007,
MNRAS, 381, 433

\bibitem[2004a]{Hubrig2004a}
Hubrig, S., Kurtz, D.~W., Bagnulo, S., et al.\ 2004a,
\aap, 415, 661

\bibitem[2004b]{Hubrig2004b}
Hubrig, S., Szeifert, T., Sch{\"o}ller, M., et al.\ 2004b,
\aap, 415, 685

\bibitem[2006]{Hubrig2006}
Hubrig, S., Briquet, M., Sch{\"o}ller, M., et al.\ 2006,
\mnras, 369, L61

\bibitem[2008]{Hubrig2008}
Hubrig, S., Sch{\"o}ller, M., Schnerr, R.~S., et al.\ 2008,
A\&A, 490, 793

%\bibitem[2009a]{hubrig09}
%Hubrig, S., Sch{\"o}ller, M., Savanov, I., et al.\ 2009a,
%AN, 330, 708

\bibitem[2009a]{Hubrig2009a}
Hubrig, S., Stelzer, B., Sch{\"o}ller, M., et al.\ 2009a,
A\&A, 502, 283

\bibitem[2009b]{Hubrig2009b}	
Hubrig, S., Briquet, M., De Cat, P., et al.\ 2009b,
AN, 330, 317

\bibitem[2010]{Hubrig2010}
Hubrig, S., Ilyin, I., \& Sch{\"o}ller, M.\ 2010,
AN, 331, 781

\bibitem[2011a]{Hubrig2011a}
Hubrig, S., Sch\"oller, M., Ilyin, I. et al.\ 2011a,
ApJL, 726, L5

\bibitem[2011b]{Hubrig2011b}
Hubrig, S., Sch\"oller, M., Kharchenko, N. V. et al.\ 2011b,
A\&A, 528, A151

\bibitem[2011c]{Hubrig2011c}
Hubrig, S., Oskinova, L.~M., Sch\"oller, M.\ 2011c, 
AN, 332, 147

\bibitem[2011d]{Hubrig2011d}
Hubrig, S., Kharchenko, N.~V., \& Sch\"oller, M.\ 2011d,
AN, 332, 65

\bibitem[2012a]{Hubrig2012a}
%Hubrig, S., Kholtygin, A., Sch\"oller, M., Langer, N., Ilyin, I., Oskinova, L.\ 2012a,
Hubrig, S., Kholtygin, A., Sch\"oller, M., et al.\ 2012a,
IBVS, 6019, 1

\bibitem[2012b]{Hubrig2012b}
Hubrig, S., Sch\"oller, M., Kholtygin, A.~F., et al.\ 2012b, 
A\&A, 546, L6

\bibitem[2012]{Ignace2012}
Ignace, R., Oskinova, L., Mass, D.\ 2012, arXiv1211.0861I

\bibitem[2000]{Ilyin2000}
Ilyin, I.\ 2000,
Ph.D.\ Thesis, University of Oulu, Finland

\bibitem[2012]{Ilyin2012}
Ilyin, I.\ 2012,
AN, 333, 213 

%\bibitem[1973]{Jones1973}
%Jones, C., Forman, W.,  Tananbaum, H., et al.\ 1973,
%ApJ, 181, L43

\bibitem[2011]{Jose2011}
Jose, J., Pandey, A.~K., Ogura, K., et al.\ 2011,
MNRAS, 411, 2530

\bibitem[1993]{Kambe1993}
Kambe, E., Ando, H., \& Hirata, R.\ 1993,
A\&A, 273, 4

\bibitem[1997]{kambe} 
Kambe, E., Hirata, R., Ando, H., et al.\ 1997,
ApJ, 481, 406

\bibitem[1996]{Kaper1996}
Kaper, L., Henrichs, H.~F., Nichols, J.~S., et al.\ 1996,
A\&AS, 116, 257 

\bibitem[1997]{Kaper1997}
Kaper, L., Henrichs, H.~F., Fullerton, A.~W., et al.\ 1997,
A\&A, 327, 281

%\bibitem[2010]{Karitskaya2010}
%Karitskaya, E.~A., Bochkarev, N.~G., Hubrig, S., et al.\ 2010,
%IBVS, 5950, 1 

\bibitem[2004]{Kharchenko2004}
Kharchenko, N.~V., Piskunov, A.~E., R{\"o}ser, S., et al.\ 2004,
AN, 325, 740

\bibitem[2005a]{Kharchenko2005} 
%Kharchenko, N.~V., Piskunov, A.~E., R{\"o}ser, S., Schilbach, E., \& Scholz, R.-D.\ 2005,
Kharchenko, N.~V., Piskunov, A.~E., R{\"o}ser, S., et al.\ 2005,
A\&A, 438, 1163 

%\bibitem[2005b]{Kharchenko2005b} 
%Kharchenko, N.~V., Piskunov, A.~E., R{\"o}ser, S., Schilbach, E., \& Scholz, R.-D.\ 2005b, A\&A, 440, 403 

\bibitem[2009]{KharchenkoRoeser2009}
Kharchenko, N.~V., \& Roeser, S.\ 2009,
``All-sky Compiled Catalogue of 2.5 million stars''

%\bibitem[2000]{LandstreetMathys2000}
%Landstreet, J.~D., \& Mathys, G.\ 2000,
%A\&A, 359, 213

\bibitem[2003]{Kholtygin2003}
Kholtygin, A.~F., Monin, D.~N., Surkov, A.~E., \& Fabrika, S.~N.\ 2003,
Astronomy Letters, 29, 175 

\bibitem[2012]{Koenig2012}
Koenigsberger, G., Moreno, E., \& Harrington, D.~M.\ 2012,
A\&A, 539, A84

\bibitem[2008]{Krey2008}
Kreykenbohm, I., Wilms, J., Kretschmar, P., et al.\ 2008,
A\&A, 492, 511 

\bibitem[2012]{Langer2012}
Langer, N.\ 2012, 
ARAA, 50, 107

\bibitem[1990]{Leon1990}
Leonard, P.~J.~T., \& Duncan, M.~J.\ 1990,
AJ, 99, 608

\bibitem[1988]{Levato1988}
Levato, H., Morrell, N., Garcia, B., \& Malaroda, S.\ 1988,
ApJS, 68, 319

\bibitem[2011]{mahy} 
Mahy, L., Gosset, E., Baudin, F., et al.\ 2011, A\&A, 525, 101

\bibitem[2004]{MaizApellaniz2004}
Ma{\'{\i}}z-Apell{\'a}niz, J., Walborn, N.~R., Galu{\'e}, H.~{\'A}., \& Wei, L.~H.\ 2004,
\apjs, 151, 103

\bibitem[2004]{Markova2004}
Markova, N., Puls, J., Repolust, T., \& Markov, H. 2004,
A\&A, 413, 693

\bibitem[2010]{Martins2010}
Martins, F., Donati, J.-F., Marcolino, W.~L.~F., et al.\ 2010,
MNRAS, 407, 1423

\bibitem[2012]{Martins2012}
Martins, F., Escolano, C., Wade, G. A., et al.\ 2012,\
A\&A, 538, 29

%\bibitem[1998]{Mason1998}
%Mason, B.~D., Gies, D.~R., Hartkopf, W.~I., et al.\ 1998,
%AJ, 115, 821
\bibitem[1993]{Mathys1993}
Mathys. G.\ 1993,
in ASP Conf.\ Ser., Vol.\ 44,
Peculiar versus Normal Phenomena in A-type and Related Stars,
ed.\ M.~M.~Dworetsky, F.~Castelli, \& R.~Faraggiana, (San Francisco: ASP), p.~232

\bibitem[1995a]{Mathys1995a}
Mathys, G.\ 1995a,
A\&A, 293, 733

\bibitem[1995b]{Mathys1995b}
Mathys, G.\ 1995b,
A\&A, 293, 746  

\bibitem[1995]{MathysHubrig1995}
Mathys, G., \& Hubrig, S.\ 1995,
A\&A, 293, 810

%\bibitem[1995c]{Mathys1995c}
%Mathys, G., \& Hubrig, S.\ 1995c,
%A\&A, 293, 810

\bibitem[2003]{Mayor2003} 
%Mayor, M., Pepe, F., Queloz, D., Bouchy, F., Rupprecht, G., et al.\ 2003,
Mayor, M., Pepe, F., Queloz, D., et al.\ 2003,
The ESO Messenger, 114, 20

\bibitem[2008]{Morel2008} 
Morel, T., Hubrig, S., \& Briquet, M.\ 2008,
A\&A, 481, 453

\bibitem[2003]{Moss2003}
Moss, D.\ 2003,
A\&A, 403, 693

\bibitem[2008]{Naze2008}
Naz{\'e}, Y., Walborn, N.~R., Rauw, G., et al.\ 2008,
\aj, 135, 1946

\bibitem[2009]{Naze2009}
Naz{\'e}, Y. \ 2009,
A\&A, 506, 1055

\bibitem[2010]{Naze2010}
Naz{\'e}, Y., ud-Doula, A., Spano, M., et al.\ 2010,
A\&A, 520, A59

\bibitem[2011]{Naze2011}
Naz{\'e}, Y., Broos, P.~S., Oskinova, L., et al.\ 2011, ApJS, 194, 7

%\bibitem[2004]{Negueruela2004}
%Negueruela, I., Steele, I.~A., \& Bernabeu, G.\ 2004,
%AN, 325, 749

%\bibitem[1997]{NoriegaCrespo1997}
%Noriega-Crespo, A., van Buren, D., \& Dgani, R.\ 1997,
%AJ, 113, 780

\bibitem[2001]{Oskinova2001}
Oskinova, L.~M., Clarke, D., Pollock, A.~M.~T.\ 2001,
A\&A, 378, L21

\bibitem[2005]{Oskinova2005}
Oskinova, L.~M.\ 2005,
MNRAS, 361, 679

\bibitem[2011]{Oskinova2011}
%Oskinova, L. M.; Todt, H.; Ignace, R.; Brown, J. C.; Cassinelli, J. P.; Hamann, W.-R., 2011, MNRAS.416.1456
Oskinova, L.~M., Todt, H., Ignace, R., et al.\ 2011,
MNRAS, 416, 1456

\bibitem[2012]{Oskinova2012}
Oskinova, L.~M., Feldmeier, A., \& Kretschmar, P.\ 2012,
MNRAS, 421, 2820

\bibitem[2013]{Petit2013}	
Petit, V., Owocki, S.~P., Wade, G.~A., et al.\ 2013,
MNRAS, {\sl in press}, also arXiv:1211.0282

%\bibitem[2010]{PflammAltenburgKroupa2010}
%Pflamm-Altenburg, J., \& Kroupa, P.\ 2010,
%MNRAS, 404, 1564

\bibitem[2012]{Pollmann2012}
Pollmann, E.\ 2012,
IBVS, 6034, 1

\bibitem[2012]{Potter2012}
Potter, A. T., Chitre, S. M., Tout, C. A.\ 2012, 
MNRAS, 424, 2358

\bibitem[1998]{Prinja1998}
Prinja, R.~K., Massa, D., Howarth, I.~D., \& Fullerton, A.~W.\ 1998,
MNRAS, 301, 926

%\bibitem[1996]{Puls1996}
%Puls, J., Kudritzki, R.-P., Herrero, A., et al.\ 1996,
%A\&A, 305, 171

\bibitem[2004]{Ramirez2004}
Ramirez S.V., Rebull L., Stauffer J., et al.\ 2004, 
AJ, 127, 2659

\bibitem[2002]{Rauw2002}
Rauw, G., Blomme, R., Waldron, W.~L., et al.\ 2002,
A\&A, 394, 993

\bibitem[2008]{Rauw2008}
Rauw, G., De Becker, M., van Winckel, H., et al.\ 2008,
A\&A, 487, 659

\bibitem[2012a]{Rauw2012a}
Rauw, G., Morel, T., \& Palate, M.\ 2012a,
A\&A, 546, A77

\bibitem[2012b]{Rauw2012b}
Rauw, G., Sana, H., Spano, M., et al.\ 2012b,
A\&A, 542, A95

\bibitem[2012]{Sana2012}
Sana, H., de Mink, S. E., de Koter, A, et al.\ 2012.
Science, 337, 444

\bibitem[2008]{SchilbachRoeser2008}
Schilbach, E., \& R{\"o}ser, S.\ 2008,
A\&A, 489, 105

\bibitem[2000]{Schulz2000}
%Schulz, N. S.; Canizares, C. R.; Huenemoerder, D.; Lee, J. C., 2000 ApJ...545L.135 
Schulz, N.~S., Canizares, C.~R., Huenemoerder, D., \& Lee, J.~C.\ 2000,
ApJ, 545, L135 

%\bibitem[1982]{SchmidtKaler1982}
%Schmidt-Kaler, Th.\ 1982,
%In: Landolt-B\"ornstein Numerical Data and Functional Relationships in Science and Technology,
%New Series, Group IV, eds.\ K.~Schafer \& H.~H.~Voigt, Springer-Verlag Press: Berlin-Heidelberg, New York
%%New Series, Group IV, eds.\ K.~Schafer \& H.~H.~Voigt, Springer-Verlag Press: Berlin-Heidelberg, New York, 2, 15

\bibitem[2011]{Snik2011} 
%Snik, F., Kochukhov, O., Piskunov, N., Rodenhuis, M., Jeffers, S., et al.\ 2011,
Snik, F., Kochukhov, O., Piskunov, N., et al.\ 2011,
%ASPC, 437, 237
in ASP Conf.\ Ser., Vol.\ 437, 
Solar Polarization 6,
ed.\ J.~R.~Kuhn, D.~M.~Harrington, H.~Lin, S.~V.~Berdyugina, J.~Trujillo-Bueno, S.~L.~Keil, \& T.~Rimmele, (San Francisco: ASP), p.~237

\bibitem[2008]{Stahl2008}
Stahl, O., Wade, G., Petit, V., et al.\ 2008,
A\&A, 487, 323

\bibitem[2008]{Sung2008}
Sung, H., Bessell, M.~S., Chun, M.-Y.,  et al.\ 2008,
AJ, 135, 441

\bibitem[2012]{Teixeira2012}
Teixeira, P.~S., Lada, C.~J., Marengo, M., \& Lada, E.~A.\ 2012,
A\&A, 540, A83

\bibitem[2011]{Townsley2011}
Townsley, L.~K., Broos, P.~S., Corcoran, M.~F., et al.\ 2011, ApJS, 194, 1 

\bibitem[1999]{Tuominen1999}
Tuominen, I., Ilyin, I., \& Petrov, P.\ 1999,
in ``Astrophysics with the NOT'', Eds.\ H.\ Karttunen \& V.\ Piirola, University of Turku, Tuorla Observatory, 47

\bibitem[2010]{Tut2010}
Tutukov, A.~V., \& Fedorova, A.~V.\ 2010,
ARep, 54, 808 

\bibitem[1988]{vanAltena1988}
van Altena, W.~F., Lee, J.~T., Lee, J.-F., et al.\ 1988,
AJ, 95, 1744

\bibitem[2006]{Loo2006}
van Loo, S., Runacres, M.~C., \& Blomme, R.\ 2006,
A\&A, 452, 1011 

\bibitem[2011]{Wade2011}
Wade, G.~A., Howarth, I.~D., Townsend, R.~H.~D., et al.\ 2011,
MNRAS, 416, 3160 

\bibitem[2012a]{Wade2012a}
Wade, G.~A., Ma\'iz Apell\'aniz, J., Martins, F., et al.\ 2012a,
MNRAS, 425, 1278

\bibitem[2012b]{Wade2012b}
Wade, G.~A., Grunhut, G., Gr\"afener, G., et al.\ 2012b,
MNRAS, 419, 2459

\bibitem[1973]{Walborn1973}
Walborn, N.~R.\ 1973,
\aj, 78, 1067

\bibitem[1982]{Walborn1982}
Walborn, N.~R.\ 1982,
\aj, 87, 1300

\bibitem[1984]{Walborn1984}
Walborn, N.~R., \& Panek, R.~J.\ 1984,
ApJ, 286, 718

\bibitem[2006]{Walborn2006}
Walborn, N.~R.\ 2006,
In: ``The Ultraviolet Universe: Stars from Birth to Death'', 26th meeting of the IAU, Joint Discussion 4, 16-17 August 2006, Prague, Czech Republic, JD04, \#19

\bibitem[2009]{Walborn2009}
Walborn, N.~R., Nichols, J.~S., Waldron, W.~L.\ 2009, ApJ, 703, 633

\bibitem[2010]{Walborn2010}
Walborn, N.~R., Sota, A., Ma{\'{\i}}z Apell{\'a}niz, J., et al.\ 2010,
ApJ, 711, L143

\bibitem[2005]{Waldron2005}
Waldron, W.~L.\ 2005,
in: R.~Smith (ed.), {\it X-ray Diagnostics of Astrophysical Plasmas: Theory, Experiment, and Observation},
AIPC, 774,  p.~353

\bibitem[2007]{Waldron2007}
Waldron, W. L., Cassinelli, J. P.\ 2007, ApJ, 668, 456

\bibitem[2005]{walker}	
Walker, G.~A.~H., Kuschnig, R., Matthews, J.~M., et al.\ 2005,
ApJ, 623, 145

%\bibitem[1957]{Zwicky1957}
%Zwicky, F.\ 1957,
%``Morphological astronomy'', Springer, Berlin

\end{thebibliography}
\end{document}